\documentclass[twocolumn,linenumbers,twocolappendix]{aastex631}

\usepackage{amsmath}
\usepackage{hyperref}
\usepackage{algpseudocode}

\definecolor{darkbrown}{HTML}{8c4600}
\definecolor{darkblue}{HTML}{1833a1}
\newcommand{\cm}{{\rm cm}}
\newcommand{\AU}{{\rm AU}}

\newcommand{\K}{{\rm K}}
\newcommand{\m}{{\rm m}}
\newcommand{\g}{{\rm g}}
\newcommand{\s}{{\rm s}}
\newcommand{\sinv}{{\rm s}^{-1}}

\newcommand{\yr}{{\rm yr}}

\newcommand{\Myr}{{\rm Myr}}

\newcommand{\Rc}{R_{\rm c}}

\newcommand{\vk}{v_{\rm k}}

\newcommand{\Sigmag}{\Sigma_{\rm g}}
\newcommand{\Sigmad}{\Sigma_{\rm d}}

\newcommand{\alphagt}{\alpha_{\rm gt}}
\newcommand{\alphavis}{\alpha_{\rm vis}}
\newcommand{\alphatot}{\alpha_{\rm tot}}

\newcommand{\Msun}{M_\odot}
\newcommand{\Mstar}{M_\star}
\newcommand{\Mdisk}{M_{\rm disk}}

\newcommand{\tcool}{t_{\rm cool}}
\newcommand{\tinfall}{t_{\rm infall}}
\newcommand{\St}{{\rm St}}

\newcommand{\cs}{c_{\rm s}}

\nolinenumbers

\submitjournal{ApJ}

\shorttitle{}
\shortauthors{Carrera, Davenport, Simon, et al.}

\begin{document}


\title{Turbulence Inhibits Planetesimal Formation in Class 0/I Disks Subject to Infall}

\author[0000-0001-6259-3575]{Daniel Carrera}
\affiliation{Department of Physics and Astronomy, Iowa State University, Ames, IA, 50010, USA}

\author{Abigail Davenport}
\affiliation{Department of Physics and Astronomy, Iowa State University, Ames, IA, 50010, USA}

\author[0000-0002-3771-8054]{Jacob B. Simon}
\affiliation{Department of Physics and Astronomy, Iowa State University, Ames, IA, 50010, USA}

\author[0000-0002-0880-8296]{Hans Baehr}
\affiliation{Department of Physics and Astronomy, The University of Georgia, Athens, GA 30602, USA}

\author{Til Birnstiel}
\affiliation{University Observatory, Faculty of Physics, Ludwig-Maximilians-Universit{\"a}t M{\"u}nchen, Scheinerstr. 1, 81679 Munich, Germany}
\affiliation{Max Planck Institute for Solar System Research, Justus-von-Liebig-Weg 3, G{\"o}ttingen, 37077, Germany}

\author[0000-0002-8138-0425]{Cassandra Hall}
\affiliation{Department of Physics and Astronomy, The University of Georgia, Athens, GA 30602, USA}

\author{David Rea}
\affiliation{Department of Physics and Astronomy, Iowa State University, Ames, IA, 50010, USA}

\author[0000-0002-1589-1796]{Sebastian Markus Stammler}
\affiliation{University Observatory, Faculty of Physics, Ludwig-Maximilians-Universit{\"a}t M{\"u}nchen, Scheinerstr. 1, 81679 Munich, Germany}


\correspondingauthor{Daniel Carrera}
\email{dcarrera@gmail.com}

\nolinenumbers

\begin{abstract}
\nolinenumbers
There is growing evidence that planet formation begins early, within the $\lesssim 1\Myr$ Class 0/I phase, when infall dominates disk dynamics.
Our goal is to determine if Class 0/I disks reach the conditions needed to form planetesimals ($\sim 100$km planet building blocks) by the streaming instability (SI). We focus on a recent suggestion that early infall causes an ``inflationary'' phase in which dust grains are advected outward.
We modified the \texttt{DustPy} code to build a 1D disk that includes dust evolution, infall, and heating and cooling sources. We ran six models and examined the implications for the SI, taking into account recent works on how the SI responds to external turbulence.
In line with other works, we find that grains are advected outward, which leads to ``advection-condensation-drift'' loop that greatly enhances the dust density at the water snowline. However, we do not see this process at the silicate line. Instead, we find a new pile up at the edge of the expanding disk. However, despite these localized enhancements, even a modest amount of turbulence ($\alpha = 10^{-3}$) leaves planetesimal formation far out of reach. The midplane dust-to-gas ratio is at least an order of magnitude below the SI threshold, even taking into account recent results on how dust coagulation boosts the SI.
For planetesimals to form in the Class 0/I phase may require a way to transport angular momentum without turbulence (e.g., disk winds) or a non-SI mechanism to form planetesimals.
\end{abstract}


%
%
\section{Introduction}
\label{sec:intro}

Planetesimals are $\sim1-1,000$ km bodies that are the building blocks of terrestrial planets and the cores of giant planets \citep{Kokubo_1996,Kokubo_2000}. As such, they are key to understanding the properties and chemical compositions of planetary bodies \citep[e.g.,][]{Chambers_2001,Morbidelli_2000,Oberg_2019}.

Despite their importance, the origin of planetesimals is arguably the largest gap in our understanding of planet formation. Theoretical models by \citet{Johansen_2007a} and \citet{Cuzzi_2008}, together with evidence from the size frequency distribution of present-day asteroids and the cratering record of Vesta \citep{Morbidelli_2009}, and the ubiquity of equal-size binaries in the Kuiper belt \citep{Nesvorny_2010,Fraser_2017} lead to a growing consensus that planetesimals likely formed through the gravitational collapse of a dense cloud of mm-cm size particles, skipping across the intermediate sizes. Among the proposed mechanisms, the streaming instability \citep[SI,][]{Youdin_2005,Johansen_2007b,Youdin_2007b} has drawn the most attention in recent years. One recent success for the SI is the reproduction of the obliquity and angular momentum distributions of trans-Neptunian binaries \citep{Nesvorny_2019,Nesvorny_2021}.

However, many crucial challenges remain for the SI. The root problem is that the conditions needed to trigger the SI \citep{Carrera_2015,Yang_2017,Li_2021,Lim_2025} seem to require larger grains and/or higher solid-to-gas ratios than predicted by dust evolution models. \citet{Carrera_2017}'s attempt to reach the conditions for the SI through disk photo-evaporation only resulted in late formation of planetesimals (too late to form planets). Pressure bumps have long been considered promising locations for the SI \citep[e.g.,][]{Lau_2022}, as they concentrate solids \citep[e.g.,][]{Taki_2016} and reduce the radial pressure gradient, which is a key parameter for the SI \citep{Sekiya_2018}. While work by \citet{Carrera_2021} showed promising results in the case of relatively large particles, a follow-up study by \citet{Carrera_2022} with smaller particles uncovered a new problem: Even if the conditions for the SI are nominally met in a pressure bump, planetesimals may still fail to form for small (possibly more realistic) particle sizes because the growth timescale of the SI may be longer than the crossing time of solids across the bump, at least in the case of pressure bumps that slow down the drifting particles without halting their drift completely.

\subsection{Streaming Instability in Class 0/I Disks?}
\label{sec:intro:class_0}

This paper is another attempt to find a realistic scenario in which a circumstellar disk triggers the SI to form planetesimals, inspired by recent work by \citet{Morbidelli_2022}, \citet{Marschall_2023}, and \citet{Estrada_2023}.

There is growing evidence that planet formation starts early. For one, Class II disks may not contain enough dust mass to reproduce the observed exoplanet population (\citealt{Manara_2018} but see \citealt{Mulders_2021} and \citealt{Liu_2024}), suggesting that, at a minimum, the process of converting dust into planetesimals occurs in the Class 0/I stage. Furthermore, the ubiquity of dust rings in ALMA images suggests that giant planet cores also form early, if indeed these rings are the result of giant planets opening gaps along their orbits. Finally, within the context of our own solar system, there is meteoritic evidence that Jupiter’s core formed within the first 1 Myr of the birth of the solar nebula \citep{Kruijer_2017,Desch_2018}.

Recently, \citet{Morbidelli_2022} suggested a scenario in which planetesimals form, via the SI, during an early ``inflationary'' stage of a Class 0/I disk. The Class 0/I stage corresponds to the time when the star and disk are shrouded by a gas envelope, and infall from the envelope plays an important role in the disk dynamics. \citet{Morbidelli_2022} argue that, due to magnetic braking, the infall is deposited close to the central star, causing the disk to spread viscously. They built a simple 1D disk with a ``two population'' model for dust evolution \citep{Birnstiel_2012}. They found that the gas outflow combines with condensation/sublimation and radial drift to produce a significant pile up of material. The proposed mechanism is that solids, as well as silicate and water vapor, are advected outward by the gas, cross the condensation line, form large solids, and then drift back in. \citet{Marschall_2023} also see this cyclical process, and additionally report a significant pile up of material.

In contrast, earlier work \citep{Birnstiel_2010} conducted 1D simulations of a protoplanetary disk with infall, following the solid-body rotation infall model of  \citet{Shu_1977}. While their models show regions where dust is advected outward, they do not report the cyclical process suggested by \citet{Morbidelli_2022}. More recently, \citet{Estrada_2023} conducted their own investigation of the SI in a 1D disk, using a model of dust growth with multiple dust species, and found a nearly opposite result: the solid concentration, and especially the particle sizes, are not nearly high enough to meet any of the proposed criteria for the SI \citep{Carrera_2015,Yang_2017,Li_2021,Lim_2025}.

Why did these works reach opposite conclusions? The most notable difference between them is that \citet{Morbidelli_2022} and \citet{Marschall_2023} have an inflationary disk, but a simple treatment of dust growth, while \citet{Estrada_2023} have a sophisticated model for dust growth, but not an inflationary disk. At the same time, \citet{Birnstiel_2010} had both, but did not explore the implications for the SI.

It is our hope that our work will help settle this issue. Our contribution is a 1D disk model that has both a sophisticated treatment of dust evolution, built on top of \texttt{DustPy} \citep{Stammler_2022}, but adopting the exact same inflationary model of \citet{Morbidelli_2022} and \citet{Marschall_2023}. In addition,

\begin{itemize}
\item We also explore a wider portion of the parameter space: We test varying stellar masses and we examine some of the simplifications in the thermal model of \citet{Marschall_2023}.

\item We added a prescription for the effect of disk self-gravity on angular momentum transport (though this turned out to be a negligible effect in the disk regions that we are interested in).

\item Importantly, we take into account recent works that explore how the SI behaves under turbulence \citep{Lim_2024} as well as a feedback mechanism between dust growth and the SI \citep{Carrera_2025}. Before \citet{Lim_2024}, the SI criteria have relied heavily on 2D simulations where the SI itself is the only source of turbulence.
\end{itemize}

This paper is organized as follows. Our experiment and numerical model are detailed in section \S\ref{sec:methods}. Results are presented in section \S\ref{sec:results}, followed by a discussion and implications in section \S\ref{sec:discussion}. Lastly, we summarize and conclude in section \S\ref{sec:conclusions}.

%
%
\section{Methods}
\label{sec:methods}

We modified \texttt{DustPy} \citep{Stammler_2022} to make a 1D model of a Class 0/I disk. \texttt{DustPy} solves the gas and dust evolution equations of a 1D axisymmetric circumstellar disk, and has been used often to model Class II disks. To adapt \texttt{DustPy} to a Class 0/I disk we added infall from the surrounding envelope, we compute the thermal evolution of the disk using the flux-limited diffusion model, and we added prescriptions for heating due to infall, turbulent viscosity, and gravitoturbulence. Since the early disk is very hot, we added the silicate condensation front in addition to the water snowline. In the rest of this section we describe each component in turn.

\subsection{Structure of a Class 0/I Disk}
\label{sec:intro:structure}

The surface density $\Sigma$ of a circumstellar accretion disk is primarily determined by its temperature $T$ and by angular momentum transport. For instance, the accretion rate of a steady-state disk is

\begin{equation}
    \dot{M} = 3\pi\nu\Sigma
\end{equation}

\noindent
where $\nu$ is the disk viscosity, which quantifies angular momentum transport. It is commonly expressed as an ``alpha viscosity'' $\nu = \alpha \cs H$ \citep{Shakura_1973} where $\cs$ is the sound speed and $H = \cs/\Omega$ is the vertical pressure scale height of the gas and $\Omega$ is the Keplerian angular frequency.

While disk self-gravity does not play a significant role in the inner disk, in the outer disk it can be an important component of angular momentum transport, can cause large-scale spiral structures, and can even lead to disk fragmentation \citep{Gammie_2001}. The strength of self-gravity is quantified by \citet{Toomre_1964}'s $Q$ stability parameter

\begin{equation}\label{eqn:Toomre_Q}
    Q \equiv \frac{\Omega\cs}{\pi G \Sigma},
\end{equation}

\noindent
where $G$ is the gravitational constant. If the disk lacks a sufficient source of heating, it will cool down until $Q \approx 1$, at which point gravitational instabilities generate spiral shocks that both heat the disk \citep{Goldreich_1965,Durisen_2007} and contribute to outward angular momentum transport \citep{Lin_1987,Laughlin_1994,Lodato_2005,Cossins_2009} allowing the disk to accrete.

For $\Mdisk < 0.5 \Mstar$ ($\Mdisk$ and $\Mstar$ are the disk and stellar masses, respectively) the disk settles into a quasi-steady state with $Q \approx 1-2$ \citep{Paardekooper_2011} where the angular momentum transport due to self-gravity can be treated as a local process with an effective $\alpha$ \citep{Lodato_2005,Rice_2016}. For a viscous disk in thermal equilibrium, \citet{Shakura_1973}'s $\alpha$ parameter can be related to the cooling time $\tcool$ via \citet{Gammie_2001}

\begin{equation}\label{eqn:alpha_cool}
    \alpha = \frac{4}{9\gamma(\gamma - 1)}(\Omega \tcool)^{-1}
\end{equation}

\noindent
where $\gamma$ is the adiabatic index and $\tcool$ is the cooling timescale.

For a more massive disk with $\Mdisk \approx \Mstar$, the disk forms a series of recurrent grand-design spirals with a dominant $m = 2$ mode. While accretion is clearly non-local in this regime, \citet{Lodato_2005} found that the effective $\alpha$ during the spiral episodes is not much larger during periods of low-spiral activity, and both values lie close to the value expected based on the balance between heating and cooling. Furthermore, the spiral structures are short-lived, as they form and dissipate on an orbital timescale. Therefore, \citet{Rice_2016} suggest that for $\Mdisk/\Mstar$ approaching unity the local approximation is still valid ``in a time-averaged sense'' 
(see also \citealt{Xu_2021} and \citealt{Bethune_2021}). In practice, our simulations never get close to $\Mdisk/\Mstar \sim 1$; our simulations peak at $\Mdisk/\Mstar < 0.15$ after $\sim 10^5 \yr$.

\subsection{The Experiment}
\label{sec:methods:experiment}

\begin{table}[ht]
\caption{Our six simulations span two initial stellar masses $M_{\star,0}$ and three infall models: A classic model \citep[][]{Shu_1977}, one where infall has a low angular momentum due to magnetic braking \citep[][]{Morbidelli_2022}, and a ``Soft Landing'' where infall additionally loses kinetic energy before the infall reaches the disk surface (this work).}
\label{tab:models}
\begin{tabular}{lll}
      & \textbf{$M_{\star,0}$} & \textbf{Infall model} \\ \hline
\texttt{Classic-M05} & 0.5 & Classic \\
\texttt{Classic-M03} & 0.3 & Classic \\
\texttt{Low-Ang-M05} & 0.5 & Low Ang Momentum \\
\texttt{Low-Ang-M03} & 0.3 & Low Ang Momentum \\
\texttt{Soft-Land-M05} & 0.5 & Soft Landing (this work) \\
\texttt{Soft-Land-M03} & 0.3 & Soft Landing (this work)
\end{tabular}
\end{table}

\begin{figure}[ht]
\centering
\includegraphics[width=0.45\textwidth]{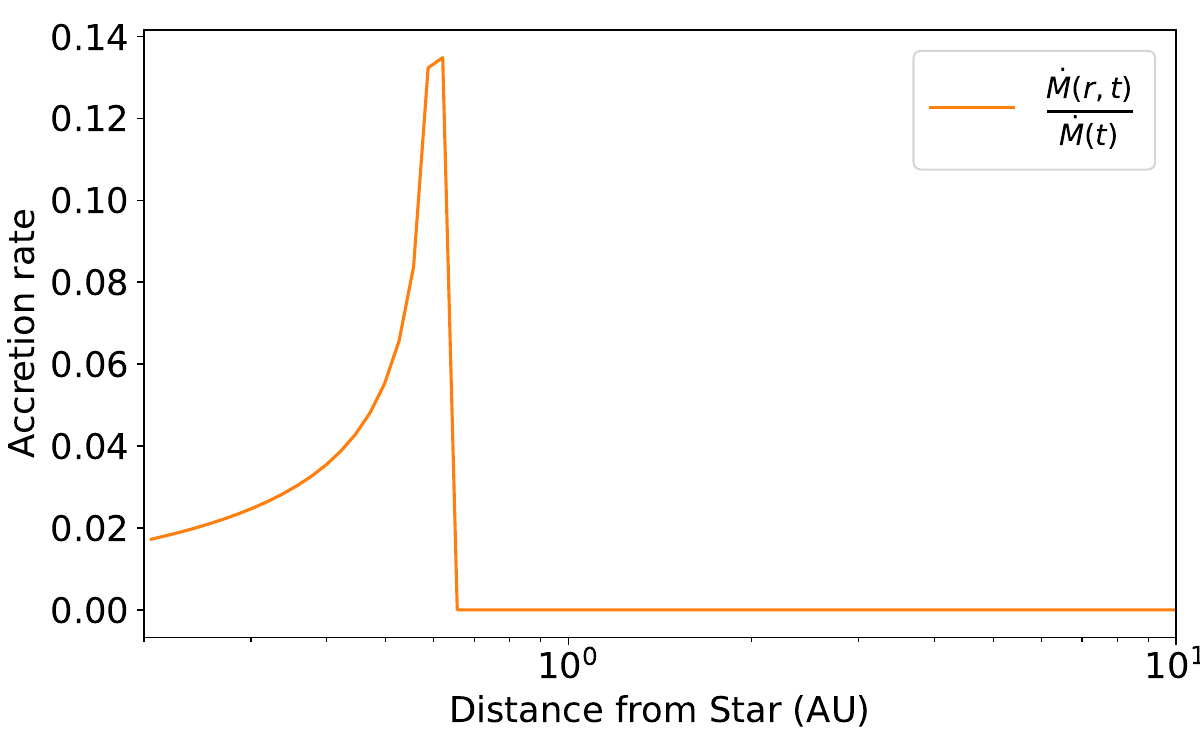}
\caption{Infall rate $\dot{M}(r,t)$ for a sample snapshot (the \texttt{Log-Ang-M03} model at $t = 10^4\yr$). The infall rate peaks at $\Rc$. See Equation \ref{eqn:mdot_r} and the main text.}\label{fig:mdot_r}
\end{figure}

Our experiment consists of three infall models and six simulations (Table \ref{tab:models}). We adopt the two infall models of \citet{Marschall_2023}, and add a third ``Soft Landing'' model where we assume that the infall loses most of its gravitational potential energy before it reaches the disk surface (see below for more details). In all cases, the total gas infall rate is

\begin{equation}
    \dot{M}(t) = \frac{M_\odot - M_{\star,0}}{\tinfall} \exp\left[-\frac{t}{\tinfall}\right]
\end{equation}

\noindent
where $M_{\star,0}$ is the initial stellar mass (either 0.5 or 0.3 $M_\odot$) and we follow \citet{Marschall_2023} in setting $\tinfall = 10^5 \yr$. Let $r_{\rm +}$ and $r_{\rm -}$ be the boundaries of a radial grid cell and let $\Rc$ be the centrifugal radius.

\begin{equation}\label{eqn:mdot_r}
    \dot{M}(r,t) = \left[
            \left(1 -  
            \sqrt{\frac{r_{\rm -}}{\Rc}}
            \right)^{1/2}
            -
            \left(1 -
            \sqrt{\frac{r_{\rm +}}{\Rc}}
            \right)^{1/2}
        \right] \dot{M}(t)
\end{equation}

\noindent
for $r_{\rm +} < \Rc$ and $0$ otherwise. Infall inside $0.05\AU$ is assumed to fall directly onto the central star, and is added to the stellar mass on every timestep. We also track the disk accretion rate and add the accreted gas mass to the star as well. To shorten the simulation time, we set the inner edge of the disk to $r_{\rm in} = 0.25\AU$ for the $M_{\star,0} = 0.5\Msun$ runs and $0.20\AU$ for the $M_{\star,0} = 0.3\Msun$ and scale $\dot{M}(r,t)$ accordingly so that $\sum_r\dot{M}(r,t) = \dot{M}(t)$. Our choices of $r_{\rm in}$ reflect the trade-off between computing speed and resolving the important regions of the disk. The runs with $M_{\star,0} = 0.3\Msun$ ran slightly faster, which allowed us to choose a smaller $r_{\rm in}$. Lastly, there is one grid cell with $r_{\rm -} < \Rc < r_{\rm +}$. That is the cell where $\dot{M}(r)$ is highest, so we were careful to account for that cell accurately (we replace $r_{\rm +}$ with $\Rc$ in Equation \ref{eqn:mdot_r}). In the end, all the infall is accounted for, and in the long run, the stellar mass converges to exactly $1 \Msun$. Figure \ref{fig:mdot_r} shows $\dot{M}(r,t)$ for a sample snapshot.

Aside from $r_{\rm in}$, our models are distinguished from each other by how we calculate $\Rc$ and how much energy is deposited onto the disk. Our infall models are:

\begin{enumerate}
\item \textbf{Classic}: Based on \citet{Shu_1977}, we model the gas cloud as a rigidly rotating sphere and is \citep{Marschall_2023}

\begin{equation}
    \Rc = 53 \AU
        \left(\frac{\omega}{10^{-14} \s^{-1}}\right)^2
        \left(\frac{T}{10\K}\right)^{-4}
        \left(\frac{M_{\rm tot}}{M_\odot}\right)^3
\end{equation}

\noindent
where $\omega = 9 \times 10^{-15}\s^{-1}$ is the angular velocity of the cloud, $T = 15\K$ is its temperature, and $M_{\rm tot}$ is the total mass of the star-disk system. The energy deposited onto the disk is

\begin{equation}\label{eqn:q_infall}
    \dot{Q}_{\rm infall} = \frac{1}{2} \frac{G M_\star(t) \dot{M}(r)}{r}
\end{equation}

\noindent
where $G$ is the gravitational constant. This model assumes that only half of the potential energy of the gas is injected into the disk and the rest is lost during the infall phase.

\item \textbf{Low Ang Momentum}: In the scenario proposed by \citet{Morbidelli_2022}, the infalling material loses angular momentum due to magnetic braking.

\begin{equation}
    \Rc = \frac{0.35 \AU}{\sqrt{M_\star(t)}}
\end{equation}

\noindent
and $\dot{Q}_{\rm infall}$ follows Equation \ref{eqn:q_infall}.

\item \textbf{Soft Landing}: Same $\Rc$ as in the \texttt{Low-Ang} model, but we additionally assume that only 1\% of the potential energy in the cloud makes its way onto the disk surface.

\begin{equation}
    \dot{Q}_{\rm infall} = 0.01 \frac{G M_\star(t) \dot{M}(r)}{r}
\end{equation}

\noindent
This model is motivated by recent 3D simulations of disk formation \citep{Mauxion_2024}; these studies indicate that the infall may lose much of its energy before landing on the surface of the disk. However, they did not follow the energy evolution in the system, nor did they resolve the inner few AU, making it difficult to quantify precisely how much energy is lost during the infall phase prior to the gas reaching the inner disk. Thus, we make a conservative estimate that 99\% of potential energy is lost before landing on the disk (e.g., through Poynting flux).

A related issue is that heating from $\dot{Q}_{\rm infall}$ occurs at the surface of the disk, not the midplane. \citet{Morbidelli_2022} use a simple relation to link surface and midplane temperatures

\begin{equation}\label{eqn:T_surf}
    T_{\rm surf}^4 = \frac{4}{3}
    \frac{2 T^4}{\kappa \Sigma}
\end{equation}

\noindent
where $\kappa$ is the opacity (we use \citet{Bell_1994} opacities). This formula is valid for an optically thick disk heated at the midplane. \citet{Morbidelli_2022}'s model computes the temperature as if all heating and cooling occurs at the midplane (see \S\ref{sec:methods:temperature}) and Equation \ref{eqn:T_surf} links $T$ and $T_{\rm surf}$, which sets the cooling rate from black-body radiation. However, $\dot{Q}_{\rm infall}$ occurs at the surface, not the midplane, and that heat has to propagate inward. As a result, Equation \ref{eqn:T_surf} may overestimate the contribution from $\dot{Q}_{\rm infall}$. The \textbf{Soft Landing} model lowers $\dot{Q}_{\rm infall}$ by 99\% in order to determine how this uncertainty impacts the model results. It is an important test for the robustness of the other models in this work.

\end{enumerate}

\subsection{Gas Evolution}
\label{sec:methods:gas}

\texttt{DustPy} uses the adiabatic sound speed

\begin{equation}
     \cs = \sqrt{\frac{\gamma k_{\rm B} T}{\mu}}   
\end{equation}

\noindent
where $\gamma$ is the adiabatic index, $k_{\rm B}$ is the Boltzmann constant, and $\mu$ is the mean molecular weight of the gas. We use $\gamma \equiv 1.4$ and $\mu \equiv 2.3 \g\,{\rm mol}^{-1}$ for the entire disk, making the approximation that $\mu$ remains constant across condensation lines.

\texttt{DustPy} solves the viscous evolution of the gas surface density $\Sigma$ via the viscous advection-diffusion equation

\begin{equation}
    \frac{\partial \Sigma}{\partial t} +
    \frac{1}{r} \frac{\partial}{\partial r} \left(r v_r \Sigma \right)
    = S_{\rm ext}
\end{equation}

\noindent
where $r$ is the radial coordinate (i.e., the distance from the star), $v_r$ is the radial velocity of the gas, and $S_{\rm ext}$ is an external source term that we use to model infall

\begin{equation}
    S_{\rm ext}(r,t) = \frac{\dot{M}(r,t)}{\pi(r_{\rm +}^2 - r_{\rm -}^2)}.
\end{equation}

\noindent
For a viscous disk, the radial velocity is given by \citep{Lynden_Bell_1974}

\begin{equation}\label{eqn:viscous_evolution_orig}
    v_r = - \frac{3}{\Sigma \sqrt{r}}
        \frac{\partial}{\partial r}
        \left(\nu \Sigma \sqrt{r}\right)
\end{equation}

\noindent
By default, \texttt{DustPy} uses the $\alpha$-prescription for disk viscosity $\nu = \alpha \cs H$ \citep{Shakura_1973}. One can express the contribution of gravitoturbulence to angular momentum transport as as an additional term $\alphagt$ (detailed in \S \ref{sec:methods:alpha_gt}).
Therefore, we write

\begin{eqnarray}
    \nu &=& \alphatot \cs H \\
    \alphatot &=& \alphavis + \alphagt,
\end{eqnarray}

\noindent
where $\alphavis \equiv 10^{-3}$ is the baseline turbulent $\alpha$. Equation~\ref{eqn:alpha_cool} ties $\alpha$ to viscous heating. In our model, viscous heating is computed from $\alphatot$. Finally, note that with our choice of infall timescale and $\alphavis$, we reach a stellar mass of $0.9 M_\odot$ and an outer radius of $\sim 50$~au  in 500~kyr. Thus, we are confident that these parameters are reasonable for a viscously accreting disk with infall.

\subsection{Disk Self-Gravity}
\label{sec:methods:alpha_gt}

\begin{figure}[ht]
\centering
\includegraphics[width=0.47\textwidth]{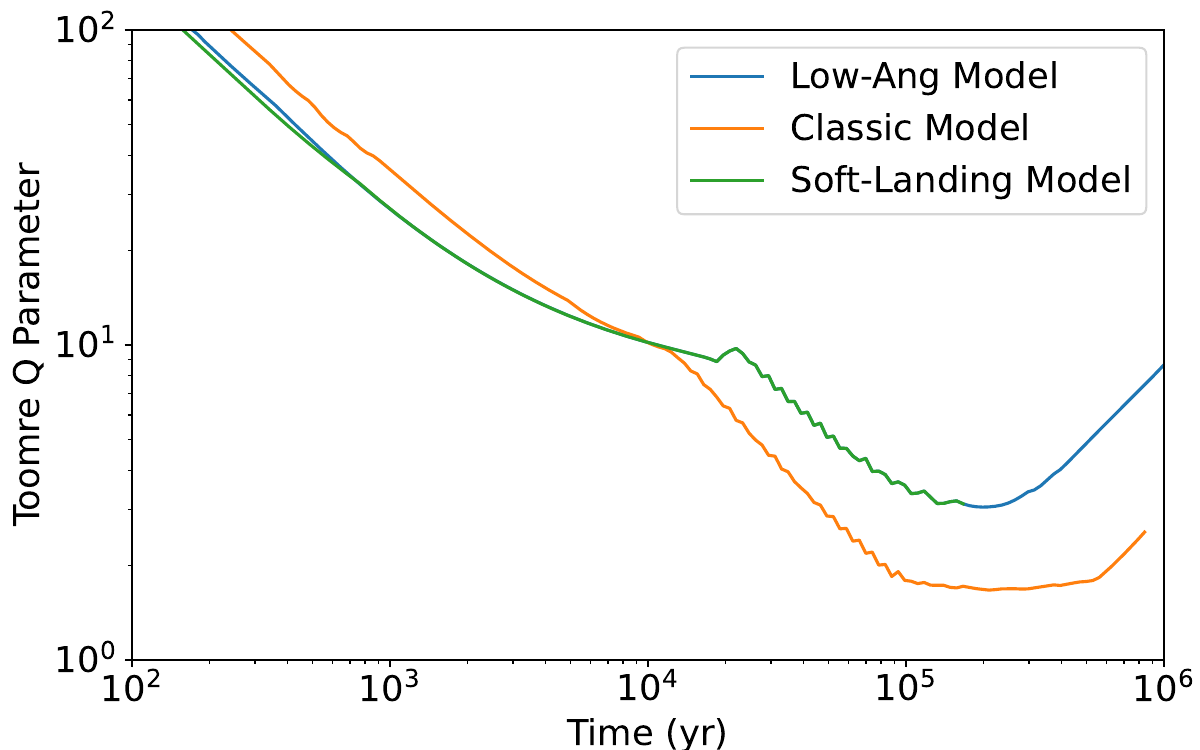}
\caption{Smallest Toomre $Q$ parameter for our \texttt{*-M05} simulations. The $Q$ values for the \texttt{*-M03} runs are similar.\label{fig:Toomre_Q}}
\end{figure}

We only consider self-gravity as a potential source of heating and angular momentum transport, and not as a source for turbulence-induced collisions between grains (as we describe below). If the disk can form spiral structures, \citet{Rice_2004} suggests that ``intermediate sized particles''' ($\St \sim 0.5 - 10$) can concentrate strongly inside. This can be safely ignored in our model because our grains are much smaller than $\St \sim 0.5$ and the disk is not massive enough to form global $m = 2$ spiral structures ($\Mdisk/\Mstar = 0.14$ for our most massive disk, which will not form such low $m$ structures; \citealt{Kratter_2016}). The model with the most massive disk is \texttt{Classic-M05}, which reaches a maximum disk mass of $\Mdisk = 0.11\Msun$ within 50 AU. This is also the least stable disk, with Toomre parameter of $Q = 1.67$. Figure \ref{fig:Toomre_Q} shows the minimum Toomre $Q$ for our \texttt{*-M05} runs --- the values for the \texttt{*-M03} runs are similar.

Our treatment of disk self-gravity follows the approach of \citet{Rafikov_2015}, who used thermal balance arguments to derive an expression for an ``effective $\alpha$'' that captures the contribution to angular momentum transport due to gravitoturbulence.

\begin{equation}\label{eqn:alphagt}
    \alphagt = \frac{8}{9}
    \frac{\sigma (\pi G Q_0)^6}{\tau + \tau^{-1}}
    \left( \frac{\mu}{k_{\rm B}} \right)^4
    \frac{\Sigma^4}{\Omega^7}
    \left(1 - \frac{T_{\rm irr}^4}{T_Q^4} \right)
\end{equation}

\noindent
where $k_{\rm B}$ is the Boltzmann constant, $Q_0 \equiv 1.5$ is the threshold for gravitational instability (GI) suggested by numerical experiments \citep{Cossins_2009,Cossins_2010}; that value is not exact \citep[e.g.][shows a density weighted $\langle Q_0 \rangle_\rho \approx 1.2$]{Bethune_2021} but we adopt $Q_0 \equiv 1.5$ to remain consistent with \citet{Rafikov_2015}. Lastly, $T_Q$ is the critical temperature for GI, which is calculated by setting $Q = Q_0$ in Equation \ref{eqn:Toomre_Q},

\begin{equation}
    T_Q = \frac{\mu}{k_{\rm B}}
    \left( \frac{\pi G Q_0}{\Omega} \right)^2 \Sigma^2.
\end{equation}

Note that Equation \ref{eqn:alphagt} gives a negative value whenever $T_{\rm irr} > T_Q$, meaning that irradiation is sufficient to bring the disk above $Q = Q_0$. When this occurs, we follow \citet{Rafikov_2015} and set $\alphagt \equiv 0$, as self-gravity does not play a role under those conditions. In any case, $\alphagt$ is a component of $\alpha$ that determines both the rate of angular momentum transport and the viscous heating of the disk.

\subsection{Dust Evolution}
\label{sec:methods:dust}

Dust evolution in \texttt{DustPy} has three components: dust transport (hydrodynamics), coagulation, and external sources. To account for transport and external sources, \texttt{DustPy} solves the advection-diffusion equation for each grain size (denoted by the subscript ``d")

\begin{equation}\label{eqn:dust_advect}
    \frac{\partial \Sigmad}{\partial t} +
    \frac{1}{r} \frac{\partial}{\partial r}
        \left[
            r v_{\rm r,d} \Sigmad
            -
            r \Sigma D_{\rm d} \frac{\partial}{\partial r}
            \left( \frac{\Sigmad}{\Sigma} \right)
        \right]
    = S_{\rm ext,d}
\end{equation}

For $S_{\rm ext,d}$, we set the total dust mass infall to

\begin{equation}
\sum_{\rm d} S_{\rm ext,d} = 0.01 S_{\rm ext}.
\end{equation}

The value of each $S_{\rm ext,d}$ is determined by the grain size distribution. After testing we found that the simulations are considerably faster if the dust grains that are falling match the local (i.e., where the infalling dust grains land) grain sizes of the disk, with minimal impact on the simulation results compared to a model in which the infalling grains follow the MRN distribution \citep{Mathis_1977}. Models with small grain infall show rapid dust growth and require very small time steps.

The $D_{\rm d}$ in Equation \ref{eqn:dust_advect} is the radial diffusivity of the dust, which is assumed to be \citep{Youdin_2007}

\begin{equation}
    D_{\rm d} = 
    \frac{\delta_{\rm r} \cs H}{1 + \St_{\rm d}^2},
\end{equation}

\noindent
where $\St_{\rm d} = t_{\rm stop} \Omega$ is the Stokes number of a grain with stopping time $t_{\rm stop}$, and $\delta_{\rm r}$ is the radial diffusion coefficient (see below) The stopping time is determined by the grain's size $a_{\rm d}$ and material density $\rho_{\rm s}$, and the local gas density. We refer the reader to the \texttt{DustPy} documentation for details of how $\St_{\rm d}$ is calculated.

Throughout all our runs we set $\delta_{\rm r} = \alphavis$, thus not including $\alphagt$. In other words, we assume that disk self-gravity can contribute to large scale angular momentum transport but that this does not translate into small-scale turbulence that can directly impact the dust. This is a valid assumption for the small ($\St_{\rm d} \le 0.1$) grains that we see in our simulations and are likely to exist in circumstellar disks, but for large ($\St_{\rm d} \ge 1$) grains gravitoturbulence leads to enhanced collision speeds \citep{Shi_2016}. Note that \texttt{DustPy} assumes isotropic turbulence. Thus, to within a factor of order unity, the turbulent diffusion parameter (i.e., turbulent Mach number squared) $\delta_{\rm turb} = \delta_{\rm r} = \delta_{\rm z}$. Based on our assumption above that $\delta_{\rm r} = \alphavis$, $\delta_{\rm turb} = \delta_{\rm r} = \delta_{\rm z} = \alphavis$. Thus, $\alphavis$ sets our dust scale height.

\begin{equation}
    H_{\rm d} = H \sqrt{\frac{\alphavis}{\alphavis + \St_{\rm d}}}.
\end{equation}

\noindent
In our model most particles are in the ``intermediate'' regime, where $t_{\rm stop}$ falls between the turnover timescale of the smallest ($t_{\rm small}$) and largest ($t_{\rm large}$) eddies. In this regime, the collision velocity is given by \citep{Ormel_2007}; again equating $\delta_{\rm turb}$ and $\alphavis$, we have

\begin{equation}\label{eqn:v_turb}
    v_{\rm d,turb} = \cs \sqrt{3 \alphavis \St_{\rm d}}.
\end{equation}

\noindent
Particles with $t_{\rm stop} < t_{\rm small}$ or $t_{\rm stop} > t_{\rm large}$ have other expressions for $v_{\rm d,turb}$ \citep{Ormel_2007}. \texttt{DustPy} takes into account not only $v_{\rm d,turb}$ but also the relative velocities due to Brownian motion and the difference in radial drift velocity for particles with different $\St$. We refer the reader to the \texttt{DustPy} documentation for more details.

\subsection{Disk Temperature}
\label{sec:methods:temperature}

We use the same method as \citet{Morbidelli_2022} to compute the disk temperature, including self-gravity $\alphagt$ as part of the viscous heating.

\begin{equation}
    \dot{Q}_{\rm visc} = 2\pi r \Delta r \frac{9}{4} \Sigma \nu \Omega^2
\end{equation}

\noindent
where $\nu = \alphatot \cs H$ and $\Delta r = r_{\rm +} - r_{\rm -}$ is the width of the grid cell. Next, we include disk cooling due to black body radiation at the surface

\begin{equation}
    \dot{Q}_{\rm bb} = 2 \times 2\pi r \Delta r \sigma_{\rm B} T_{\rm surf}^4
\end{equation}

\noindent
where $\sigma_{\rm B}$ is the Stephan-Boltzmann constant, and $T_{\rm surf}$ is the surface temperature, which is tied to the midplane temperature $T$ by Equation \ref{eqn:T_surf}.

Lastly, we include the energy flux between adjacent rings using a flux-limited diffusion model. A ring gains or loses energy at a rate $\Delta F = F_{\rm +} - F_{\rm -}$, where $F_{\rm +}$ and $F_{\rm -}$ are the energy fluxes across the exterior and interior cell boundaries respectively.

\begin{equation}
    F = (2\pi)^{3/2} \frac{16\lambda \sigma_{\rm B}}{\kappa \rho}
    \frac{d T}{d r} T^3 r H
\end{equation}

\noindent
where $\rho = \Sigma / (H \sqrt{2\pi})$ is the volume gas density at the midplane and $\lambda$ is the flux limiter, which we compute using the method of \citet{Bitsch_2013}

\begin{equation}
    \lambda =  \left\{
    \begin{array}{cc} 
        2 / (3+\sqrt{9+10 R^2})
        \quad &  \mbox{for} \quad  R \leq 2  \\
        10 / (10R + 9 + \sqrt{81+180 R}) \quad & \mbox{for} \quad R > 2 
    \end{array}
    \right.
\end{equation}

\noindent
where

\begin{equation}
 R = \frac{1}{\rho \kappa} \frac{|\nabla E_R|}{E_R}.
\end{equation}

\noindent
and $E_R = 4\sigma_{\rm B} T^4/c$ is the radiative energy density in the optically thick limit. To estimate $\kappa, E_R, \rho, T, H$ at the boundaries between adjacent rings, we compute the arithmetic average of their values inside the rings.

With these ingredients, we can compute the change of internal energy $\Delta Q$ of a ring over a timestep $\Delta t$

\begin{equation}
    \Delta Q = \left(
        \dot{Q}_{\rm infall} + \dot{Q}_{\rm visc} - \dot{Q}_{\rm bb} + \Delta F
    \right) \Delta t.
\end{equation}

\noindent
Then the change in temperature is

\begin{equation}
    \Delta T = \frac{\Delta Q}{C_{\rm v} m_{\rm ring}}
\end{equation}

\noindent
where $m_{\rm ring} = \pi(r_{\rm +}^2 - r_{\rm -}^2)(\Sigma + \Sigmad)$ is the mass of the ring,

\begin{equation}
    C_{\rm v} = \frac{\mathcal{R}}{(\gamma - 1)\mu}
\end{equation}

\noindent
is the ring's heat capacity and $\mathcal{R}$ is the gas constant. The stellar irradiation plays an indirect role in that we also assume that the disk temperature cannot fall below that of a passively irradiate disk $T_{\rm irr} = 115\K (r/\AU)^{-3/7}$.

\subsection{Snowlines}

While the molecular weight of the gas $\mu$ is kept fixed across evaporation fronts, we do model how the sticking properties of silicate and icy grains affect their sizes. The standard way to implement snowlines in \texttt{DustPy} is by altering the solid fragmentation speed $v_{\rm frag}$ across a temperature threshold; collision speeds below $v_{\rm frag}$ are assumed to lead to grains sticking while those above $v_{\rm frag}$ lead to fragmentation.

We start silicate evaporation at $T = 1,000\K$. \texttt{DustPy} does not have a mechanism to track silicate vapor. Thus, to simulate silicate vapor we set $v_{\rm frag}$ to an arbitrary small value ($v_{\rm frag} = 10\,\cm\,\sinv$) to keep grains small and well-coupled to the gas.

\begin{equation}
    v_{\rm frag} =  \left\{
    \begin{array}{rccl}
        10\,\cm\,\sinv \quad & \mbox{for} &        & T  >   1,000\K\\
         1\,\m \,\sinv \quad & \mbox{for} & 170\K <& T \leq 1,000\K\\
        10\,\m \,\sinv \quad & \mbox{for} &  95\K <& T \leq 170\K  \\
         1\,\m \,\sinv \quad & \mbox{for} &        & T \leq 95\K
    \end{array}
    \right.
\end{equation}

The collision velocity that leads to sticking for SiO$_2$ grains is highly dependent on grain size and morphology \citep{Poppe_2000}. Here we adopt $v_{\rm frag} = 1\,\m \,\sinv$ for silicate grains; this value is commonly used in the literature \citep[e.g.,][]{Carrera_2017} and is the sticking threshold
for $\sim 1 \mu\m$ spherical grains \citep{Poppe_2000,Blum_2008}.

The sticking behavior of icy grains is less well understood and more complex than silicate grains. Simulations of ice-dust aggregates by \citet{Wada_2009} suggest $v_{\rm frag} \sim 50\,\m\,\sinv$, but \citet{Gundlach_2015} conducted lab experiments of icy grains that suggest a value closer to $10\,\m\,\sinv$. In this investigation we opt for the latter value as the conservative choice. The next complication is that the sticking force is temperature dependent. \citet{Musiolik_2019} measured the rolling and sticking forces of water and found that just $25\K$ below the condensation line the forces decrease by over an order of magnitude, reducing the sticking velocity by a factor of 32. Then, a follow-up numerical study by \citet{Musiolik_2021} suggested that UV irradiation produces a liquid-like shell on icy grains, which may damp collisions. They predict that a shell just a few microns thick would be enough to allow the growth of cm-size clusters of grains.

In an effort to capture the essence of these results, we opted to maintain $v_{\rm frag} = 10\,\m\,\sinv$ for icy grains over a temperature interval of $75\K$ from the snowline and then lower it to $v_{\rm frag} = 1\,\m\,\sinv$.

\subsubsection{Sidenote On Silicate Vapor}

As noted above, \texttt{DustPy} does not have a mechanism for tracking multiple vapor species. To mimic the behavior of silicate vapor, we arbitrarily set $v_{\rm frag} = 10\,\cm\,\sinv$ for $T > 1,000\K$. This makes the silicate grains small and well-coupled to the gas, so that they follow the flow of the gas, just as silicate vapor would, while allowing us to keep track of the silicate mass. When these small grains are advected outward, or the disk cools below $T = 1,000\K$, the increase in $v_{\rm frag}$ mimics silicate condensation. There are two ways that small grains differ from silicate vapor: They are less coupled to the gas, and when they cross the silicate condensation line, instead of condensing (as vapor would) they have to grow gradually by sticking collisions. To quantify how this method may affect our results, we re-ran one of our simulations (\texttt{Low-Ang-M05}) with $v_{\rm frag} = 0.1\,\cm\,\sinv$. We found that the simulation results are identical (except for the ``grain size'' inside the silicate line) albeit with higher computational cost due to smaller time steps.

In the next section we only present the plots from our initial set of runs, with the knowledge that the larger-than-intended grain sizes for $T > 1,000\K$ do not impact our results.

%
%
\section{Results}
\label{sec:results}

\subsection{Inflationary Disk and Infall}
\label{sec:results:intro}

\begin{figure*}
\centering
\includegraphics[width=1\textwidth]{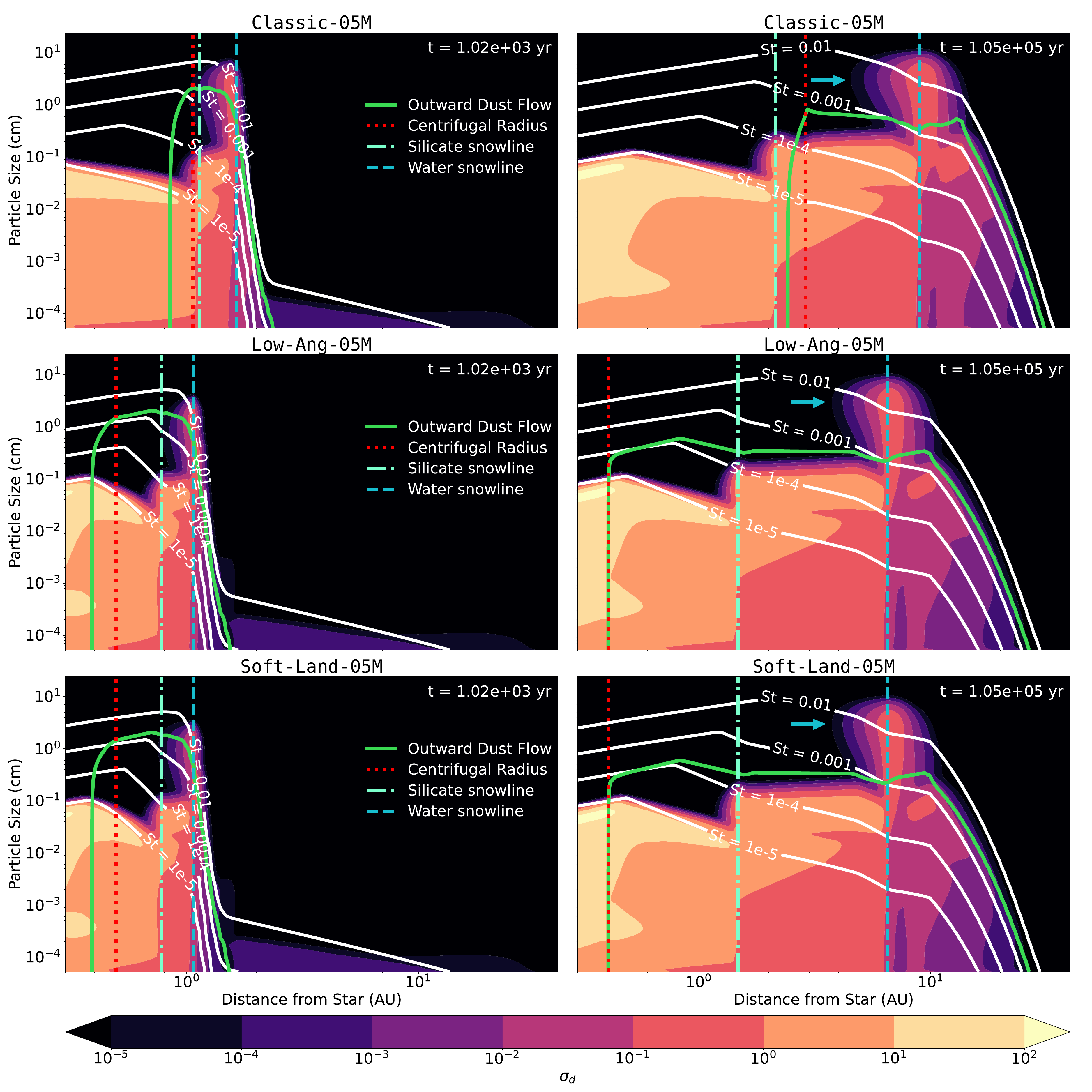}
\caption{Snapshots of the surface density of dust grains ($\Sigmad$) at $10^3$ yr (left) and $10^5$ yr (right) for all models with $M_{\star,0}~=~0.5\Msun$. The green contour marks the range of distances and grain sizes where the net radial velocity of the dust is outward. Infall occurs up to the centrifugal radius (red dotted line). The silicate ($T = 1,000\K$) and water ($T = 170\K$) snowlines are marked as cyan dot-dashed and blue dashed lines, respectively, as well as contours for Stokes numbers $\St = 10^{-2}$ and $10^{-3}$ (white). The blue arrows at $10^5$ yr show a feature where small dust grains are advected outward, cross the water snowline, and then drift back inward as they grow in size at the snowline. This loop causes a small increase in surface density at or just interior to the water snowline (see Figure \ref{fig:sigma}).\label{fig:dust_density}}
\end{figure*}

Figure \ref{fig:dust_density}  shows snapshots at $t = 10^3$ and $10^5$ yr of the dust density profile for each of the models with $M_{\star,0} = 0.5\Msun$. In all cases we see an expanding disk as it spreads viscously. The \texttt{Classic-M05} model, with its larger angular momentum and centrifugal radius, extends further out.

In contrast, the difference between \texttt{Low-Ang-M05} and \texttt{Soft-Land-M05} is barely discernible at $10^3$ yr (the latter has slightly more cm-size grains at $\sim 1$ AU), and the models are nearly indistinguishable at $10^5$ yr. We find that the temperature profile (discussed in Section \ref{sec:results:SPD_T}) is nearly identical in the two models. Even if we focus on just the heating sources and the earliest times ($10^3$ yr) we find that $Q_{\rm infall}$ only dominates the heating inside 0.6 AU for \texttt{Low-Ang} and in a narrow band between 0.30 and 0.45 AU for \texttt{Soft-Land}. In other words, $Q_{\rm infall}$ does not dominate the physics.

\citet{Mauxion_2024} showed that infall loses much of its energy before it even reaches the disk surface. When it does, the shock heating at the disk surface needs to propagate down toward the midplane. The detailed physics of this process are beyond the scope of this work. Instead, we simply consider two extreme cases:

\begin{itemize}
\item \textbf{\texttt{Low-Ang} model:} Assume that 50\% of the gravitational potential energy reaches the midplane.

\item \textbf{\texttt{Soft-Land} model:} Assume that only 1\% of the energy reaches the midplane.
\end{itemize}

\begin{figure}[ht]
\centering
\includegraphics[width=0.45\textwidth]{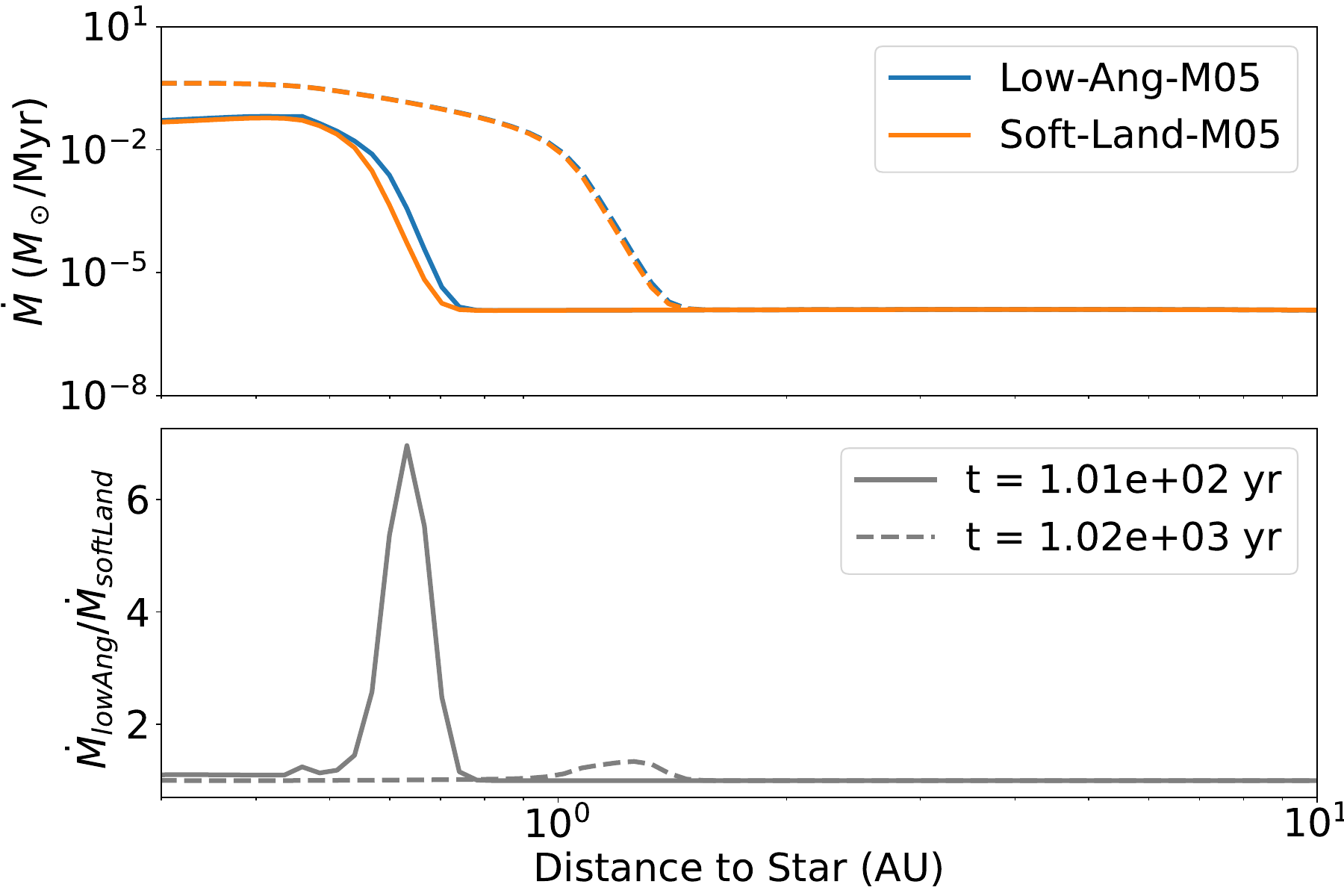}
\caption{\textit{Top:} Accretion rate (in units of $M_\odot$ per Myr) of the \texttt{Low-Ang-M05} and \texttt{Soft-Land-M05} models at $10^2$ yr and $10^3$ yr. \textit{Bottom:} Ratio of their accretion rates. At $10^2\yr$ there is a significant difference right at the edge of the disk. At $10^3\yr$ there is a small but detectable difference, again at the edge of the disk. At $10^4 \yr$ (not shown) the accretion rates are indistinguishable\label{fig:mdot}}
\end{figure}

\noindent
The true behavior of infall must lie between these extremes. Figure \ref{fig:mdot} shows the accretion rate $\dot{M}$ for these two models. It shows that $Q_{\rm infall}$ has a minimal effect on the physics. We do see some positive feedback where $Q_{\rm infall}$ increases the temperature, which adds viscosity since $\nu \propto \cs^2 \propto T$) thus enhancing viscous heating. However, the ultimate effect on both $T$ and $\nu$ is generally minor. The most significant difference between \texttt{Low-Ang} and \texttt{Soft-Land} is that the latter has a lower accretion rate in a narrow region at the edge of the disk. However, this feature is short lived: It is prominent at $10^2$ yr and visible $10^3$ yr (Figure \ref{fig:mdot}), but by $10^4$ yr (not shown in the figure) there is essentially no difference at all.

\subsection{Solid Pileup at the Disk Edge}
\label{sec:results:disk_edge}

\texttt{DustPy} computes the radial speed of dust particles, taking into account the gas viscous evolution and the inward radial drift of solids relative to the gas

\begin{eqnarray}\label{eqn:v_dust}
    v_{r,\rm dust} &=& \frac{v_{r,\rm gas} -2\eta\vk}{1 + \St^2}\\
    \eta &=& -\frac{1}{2} \left(\frac{\cs}{\vk}\right)^2
            \frac{\partial \ln P}{\partial \ln r}
\end{eqnarray}

\noindent
\texttt{DustPy} can optionally include back-reaction, but we did not explore that effect in our model. We estimate that the error in $v_{r,\rm dust}$ is no more than $\sim 2\%$ for our runs. The green contour of Figure \ref{fig:dust_density} marks the region where dust moves outward ($v_{r,\rm dust} > 0$).

In a Class II disk $v_{r,\rm gas}$ is small and often neglected, but in a rapidly expanding disk it can dominate and lead to extended regions where dust flows outward. As distance from the star increases, there is a point where the $2\eta\vk$ term in Equation \ref{eqn:v_dust} dominates over $v_{r,\rm gas}$ and $v_{r,\rm dust}$ changes sign. Because $v_{r,\rm dust}$ depends on the grain size, the extent of the region with dust outflow is also grain size dependent. At $10^3$ yr all models show a sharp edge where all grains stop their outward flow at nearly the same location. That edge softens as the gas continues to expand viscously, preferentially carrying smaller grains with it.

\begin{figure}[ht]
\centering
\includegraphics[width=0.45\textwidth]{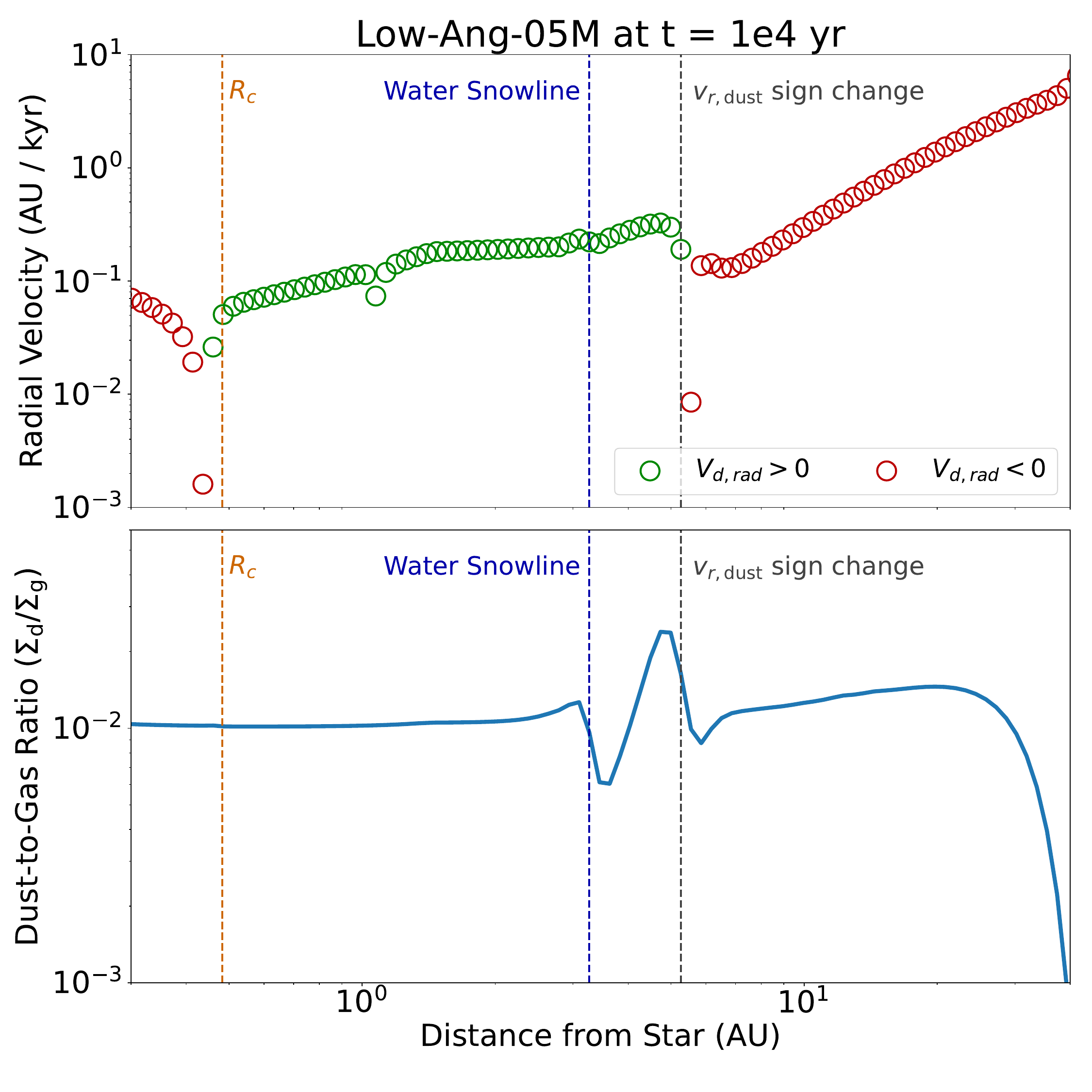}
\caption{\textit{Top:} Mean (mass-weighted) radial velocity of dust, $v_{r,\rm dust} = \sum \rho_{\rm i} v_{\rm i} / \sum \rho_{\rm i}$, for the \texttt{Low-Ang-M05} model at $10^4\yr$. The dashed lines show the location of the centrifugal radius $R_{\rm c}$, the water snowline, and where $v_{r,\rm dust}$ changes sign. \textit{Bottom:} Dust-to-gas ratio for the same snapshot. The plot shows two distinct dust pileups: One associated with the water snowline, and another with the change in sign of $v_{r,\rm dust}$.}\label{fig:V_rad}
\end{figure}

Figure \ref{fig:V_rad} shows the radial velocity of particles and the dust-to-gas ratio at $10^4\yr$. We choose $t = 10^4\yr$ because this is the snapshot that most clearly shows that there are two distinct dust pile-ups: one associated with the snowline, and a different one associated with the transition between $v_{r,\rm dust} > 0$ to $v_{r,\rm dust} < 0$, at the edge of the expanding disk. At $t = 10^3\yr$ the two pileups mostly overlap and are difficult to distinguish, and at $t = 10^5 \yr$ the outer pileup has largely dissipated. Furthermore, the bottom of Figure \ref{fig:V_rad} shows that, at $10^4\yr$, an inflationary disk can double the dust-to-gas ratio at the outer edge: The typical dust-to-gas ratio across the disk is $Z = 0.01$ but close to the point where $v_{r,\rm dust}$ changes sign, it reaches $Z = 0.024$, while the pile up just interior to the snowline is small. 

\begin{figure*}[h!]
\centering
\includegraphics[width=1\textwidth]{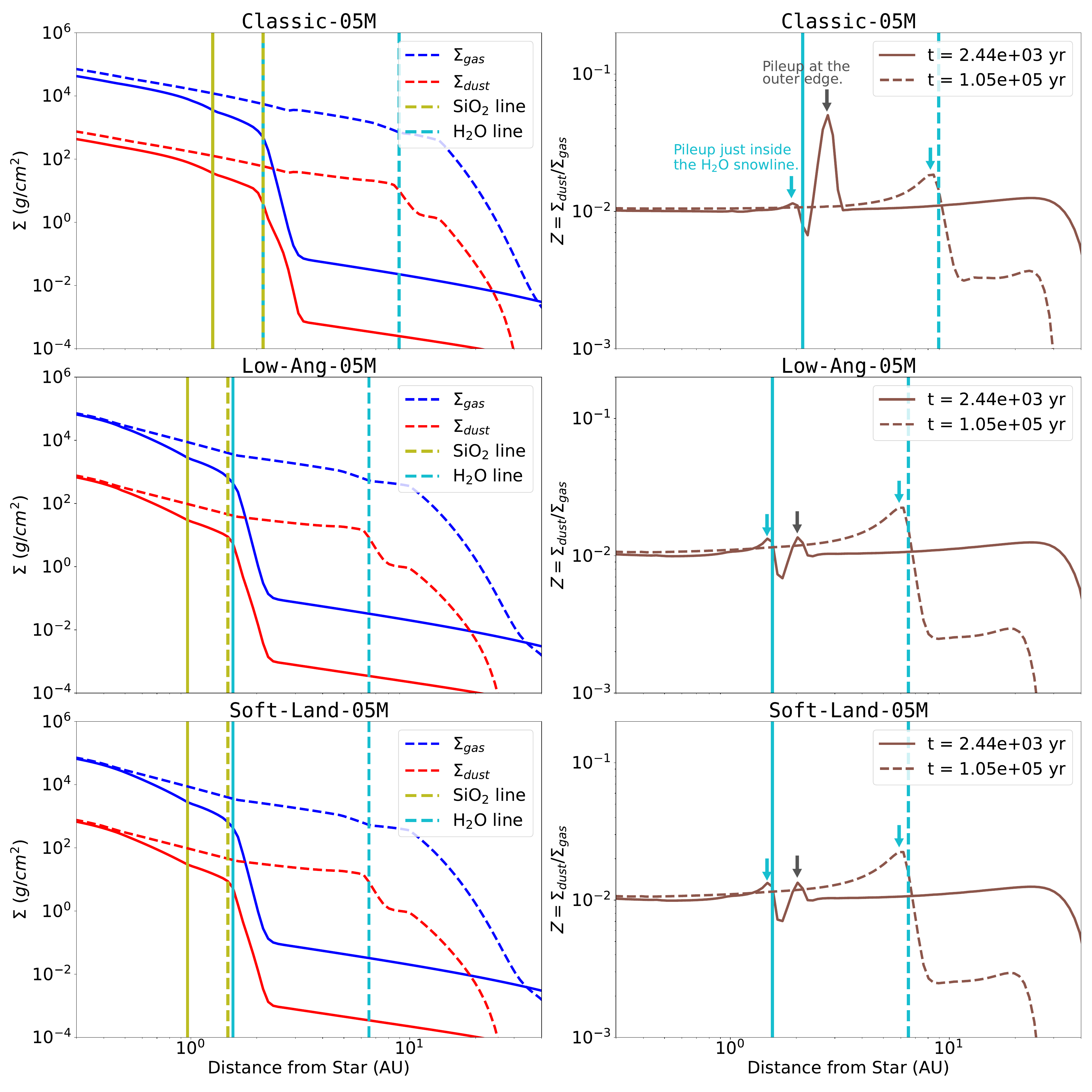}
\caption{Snapshots of $\Sigma_{\rm gas}$ and $\Sigma_{\rm dust}$ at $2.44\times10^3\yr$ and $10^5$ yr for all models with an initial stellar mass of $0.5\Msun$. The $2.44\times10^3\yr$ was chosen because it is the earliest snapshot that shows a dust pileup at the outer edge. \textit{Left:} Gas and dust density. \textit{Right:} Dust-to-gas ratio. The silicate ($T = 1,000\K$) and water ($T = 170\K$) snowlines are marked. The blue arrows point to the pileup just interior to the water snowline. The gray arrows point to the pileup associated with the outer edge of the disk, which appears at $2.44\times10^3\yr$ and dissipates by $10^5\yr$.\label{fig:sigma}}
\end{figure*}

Figure \ref{fig:sigma} shows the surface density of the gas and dust, as well as the dust-to-gas ratio $Z = \Sigmad/\Sigmag$ for all three models at $10^3\yr$ and $10^5\yr$. At $10^5\yr$, the pile up at the snowline has grown to almost double the initial dust-to-gas ratio, while the pileup at the edge of the disk has disappeared.

\subsection{Limited Pileup at Snowlines}
\label{sec:results:snowlines}

The key result of \citet{Morbidelli_2022} is that the water and silicate condensation fronts both produce a large pileup of solids that can lead to contemporaneous planetesimal formation at both locations. This is caused by an ``advect-grow-drift'' cycle in which silicate and water vapor flow outward by advection and diffusion, which promotes grain growth, and then those grains experience radial drift so that they cross the sublimation line once again.

We find no evidence of the ``advect-grow-drift'' cycle or a solid pile up at the silicate line. While our runs do not explicitly track silicate vapor, our $v_{\rm frag}$ transition at the silicate line should mimic silicate vapor very accurately. In effect, we model silicate vapor as $\St \sim 10^{-5}$ grains, which are nearly perfectly coupled to the gas and experience advection and diffusion just as silicate vapor would. The green contours in Figure \ref{fig:dust_density} show the region where solid grains move outward ($v_{r,\rm dust} > 0$). Notice that the grains on the far side of the silicate line are still inside the $v_{r,\rm dust} > 0$ contour. For this reason, the ``advect-grow-drift'' cycle cannot occur.

For completeness, we looked at every snapshot in our simulations. We did find a brief moment at $0.2\Myr$ where the \texttt{Low-Ang-M05} and \texttt{Soft-Land-M05} models do show larger grains turning around at the silicate line. However, this occurs only for a brief moment, and has no discernible effect on the dust density.

Would our results change if we modeled silicate condensation? We argue that that they would not: The grain sizes near the silicate line are limited by the fragmentation barrier. Condensation may increase the grain growth rate, but once grains reach the fragmentation barrier, collisions between grains halts further growth. Therefore, the key finding that silicate grains are too small to drift back in would remain true. 

We do find clear evidence of the ``advect-grow-drift'' cycle at the water snowline, as seen in Figure \ref{fig:dust_density}. The green contours mark the region where $v_{r,\rm dust} > 0$. Across the water snowline, grains grow large enough to exit the green contour. The blue arrows in Figure \ref{fig:dust_density} show the large grains from the water snowline drifting back in. We also see, in Figure \ref{fig:sigma} (cyan arrows), that just inside the water snowline there is a clear accumulation of solids. 

These results are in partial agreement with \citet{Morbidelli_2022} in that they also found a ``advect-grow-drift" cycle at the water snowline. However, we do not see this behavior at the silicate line, whereas they did. We speculate on the reasons for the differences in Section~\ref{sec:discussion:estrada}

\subsection{CAIs and Snowlines}
\label{sec:results:SPD_T}

\begin{figure}[ht]
\centering
\includegraphics[width=0.45\textwidth]{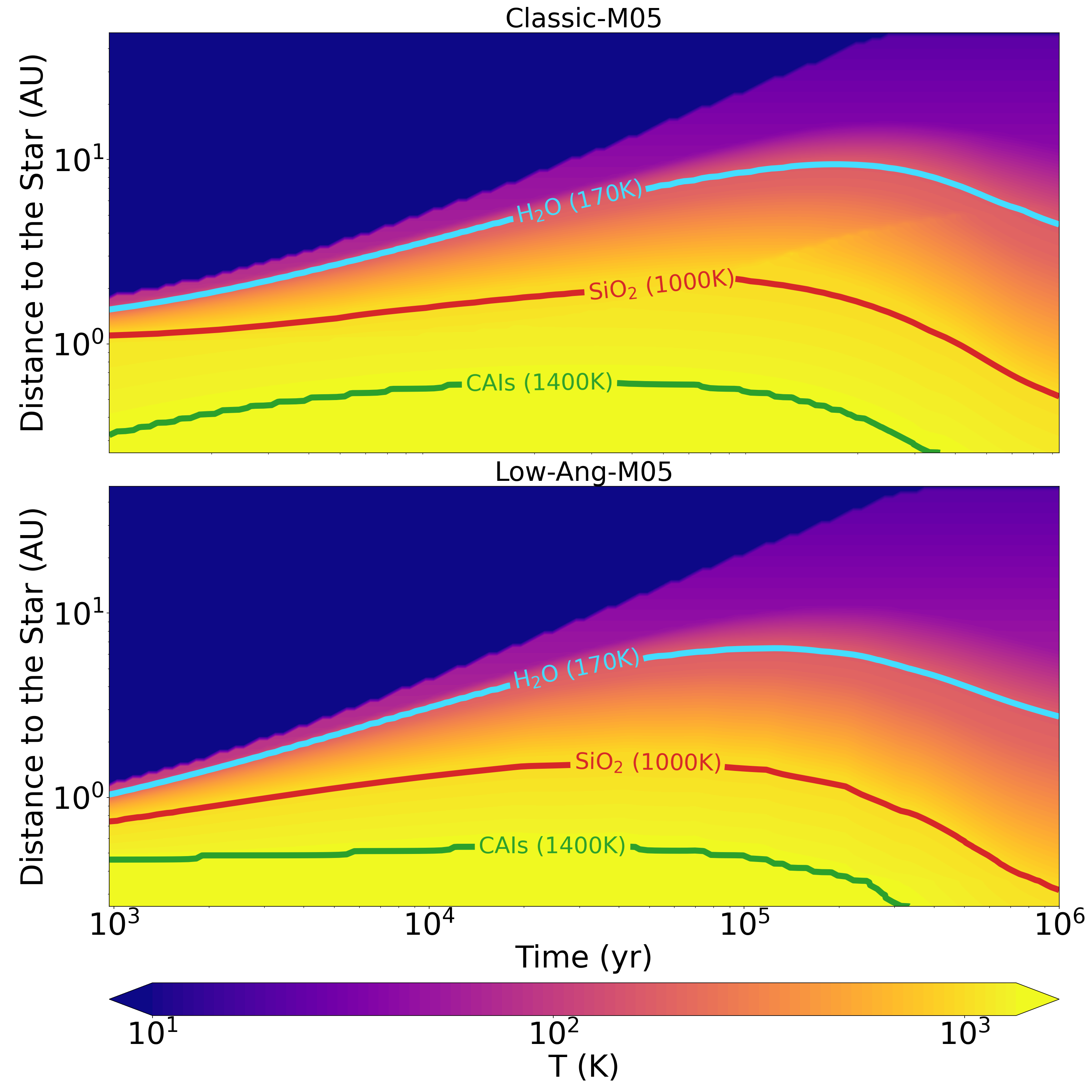}
\caption{Spacetime diagrams of the disk temperature for the \texttt{Classic-M05} and \texttt{Low-Ang-M05} models. We also show the water snowline (cyan), the silicate condensation line (red) and the temperature where Ca-Al-rich Inclusions (CAIs) probably condense (green).\label{fig:SPD_T}}
\end{figure}

Figure \ref{fig:SPD_T} shows the disk temperature profile of the \texttt{Classic-M05} and \texttt{Low-Ang-M05} models over the entire $1\Myr$ simulation. We do not show \texttt{Soft-Land-M05} as it is nearly the same as \texttt{Low-Ang-M05}.

In Figure \ref{fig:SPD_T} we intentionally remove the outer region where the disk has not yet reached through viscous spreading. For the purpose of these figures, we define the outer edge of the disk as the location where $\Sigmag$ has increased in time by a factor of 100 from its initial value $\Sigmag(t=0)$ as the disk expands outward. We found this to be a reliable way to identify the exponential drop-off in $\Sigmag$ at the outer edge.

One of the most important problems in meteoritics is the formation of Calcium-Aluminum–rich Inclusions (CAIs). Most CAIs formed as fine-grained condensates at very high temperatures ($> 1,300\K$). Figure \ref{fig:SPD_T} shows a contour line at $T = 1,400\K$ as a rough estimate of the time and place where CAIs might have condensed. Our results suggest that, regardless of the infall model, CAIs can form in the inner disk, as far out as $\sim 0.5\AU$, throughout the first $\sim 0.3\Myr$ of the disk's life. This timing remains the same in our \texttt{*-M03} models, and is remarkably consistent with what is known of the chronology of CAI formation, which appear to have formed in a $\sim 0.16\Myr$ window \citep{Connelly_2012}. Interestingly, the location the CAI formation line is relatively constant for the first $10^5$ years of the \texttt{Low-Ang-M05} simulation; unfortunately this is not a testable prediction as CAIs spread over a wide region through advection, drift, and diffusion \citep{Woitke_2024}.

In the last section we noted an increase in the solid-to-gas ratio around the water snowline. Considering that the water snowline may play an important role in planet formation, it is notable that it sweeps through the disk, starting at $\sim 1\AU$ and then reaches as far as $\sim 6\AU$ at around $\sim 0.2\Myr$ before moving back toward the star.

\subsection{Grain Sizes and Planetesimal Formation}
\label{sec:results:planetesimal_formation}

\begin{figure}[!t]
\centering
\includegraphics[width=0.45\textwidth]{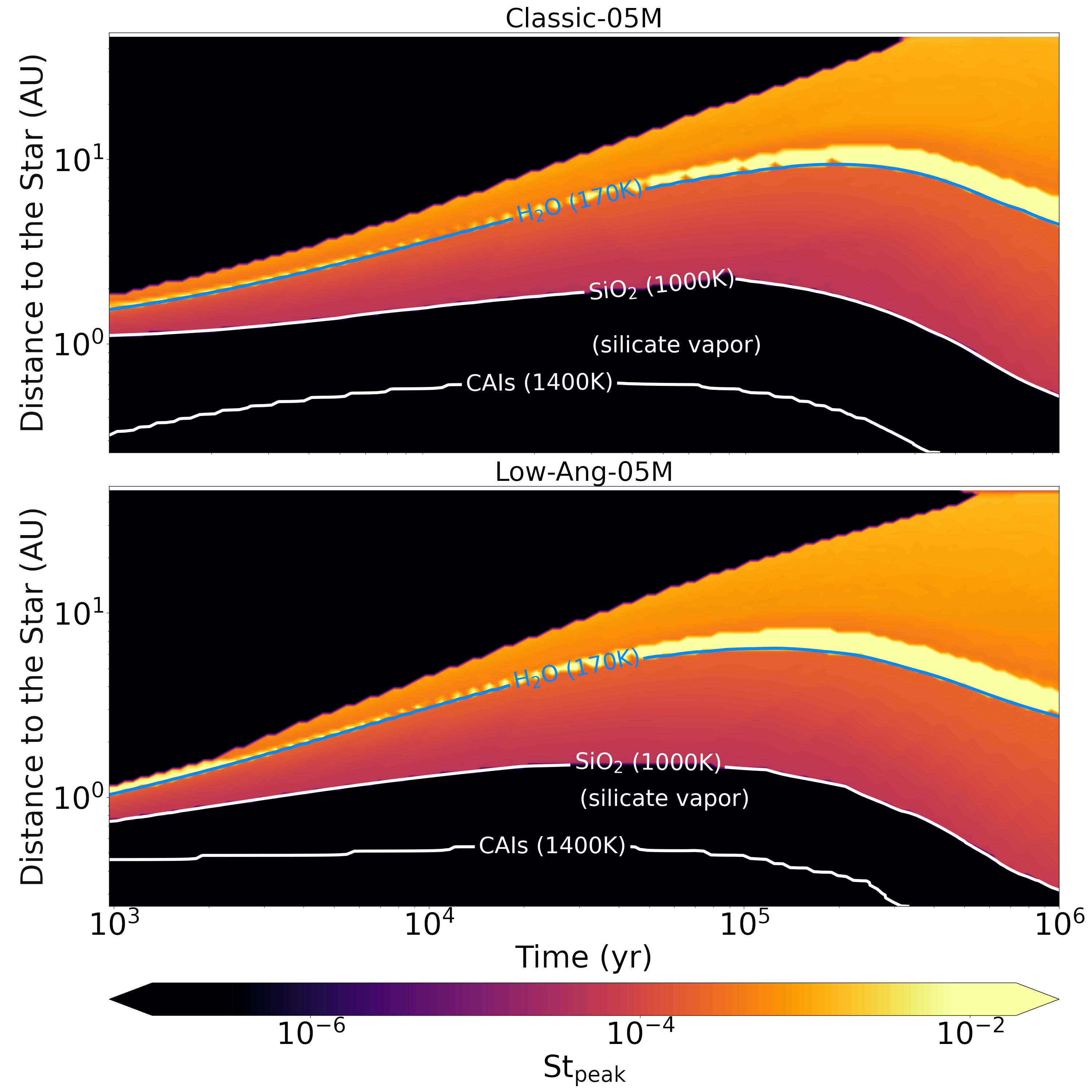}
\caption{Spacetime diagrams of $\St_{\rm peak}$ --- the particle Stokes number where most of the mass is concentrated --- for the \texttt{Classic-M05} and \texttt{Low-Ang-M05} models. There are three sharp transitions in particle size, corresponding to the silicate line, the water snowline, and the point at lower temperatures where water ice becomes less sticky \citep{Musiolik_2019}.\label{fig:SPD_St}}
\end{figure}

\begin{figure*}[!h]
\centering
\includegraphics[width=1\textwidth]{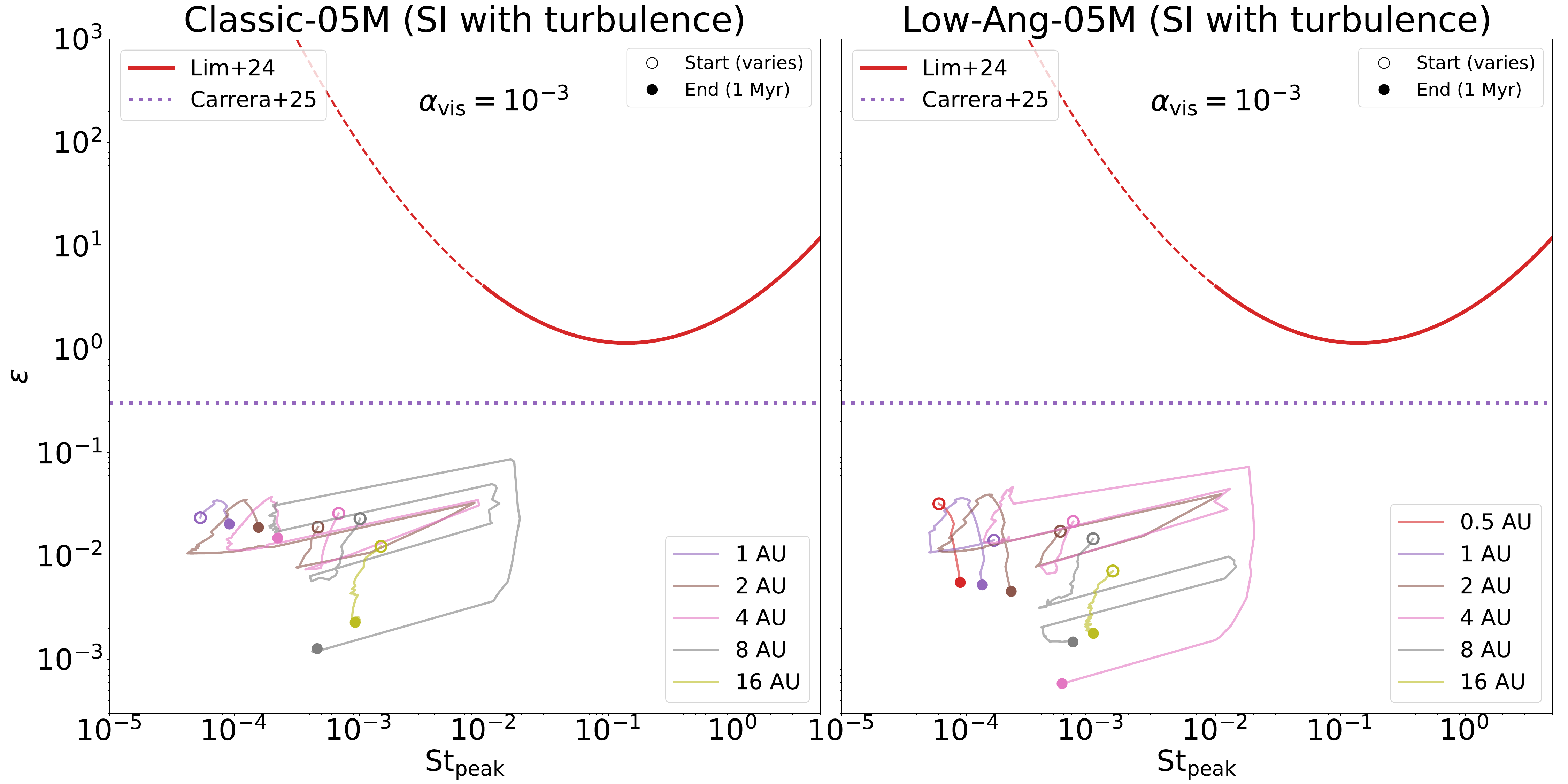}
\caption{Particle tracks at select locations in the disk for the \texttt{Classic-M05} and \texttt{Low-Ang-M05} models. The red curve is the SI criterion of \citet{Lim_2024}, which accounts for how turbulence affects the SI (the dashed line is an extrapolation of their result). The purple dotted line is the threshold of \citet{Carrera_2025}, where the combination of dust coagulation and turbulence dampening may be able to bring SI filaments into the planetesimal-forming region. We use $\St_{\rm peak}$ (Equation \ref{eqn:St_peak}) as the characteristic Stokes number. The particle tracks start at the time when $T < 1,000\K$ (i.e., when silicates condense) and/or the disk edge reaches that radial location. The key result is that, even accounting for the \textit{concentration-coagulation} feedback loop of \citet{Carrera_2025}, the conditions of the disk do not allow planetesimal formation via the SI. \label{fig:eps_vs_St_05M}}
\end{figure*}

\begin{figure*}[!h]
\centering
\includegraphics[width=1\textwidth]{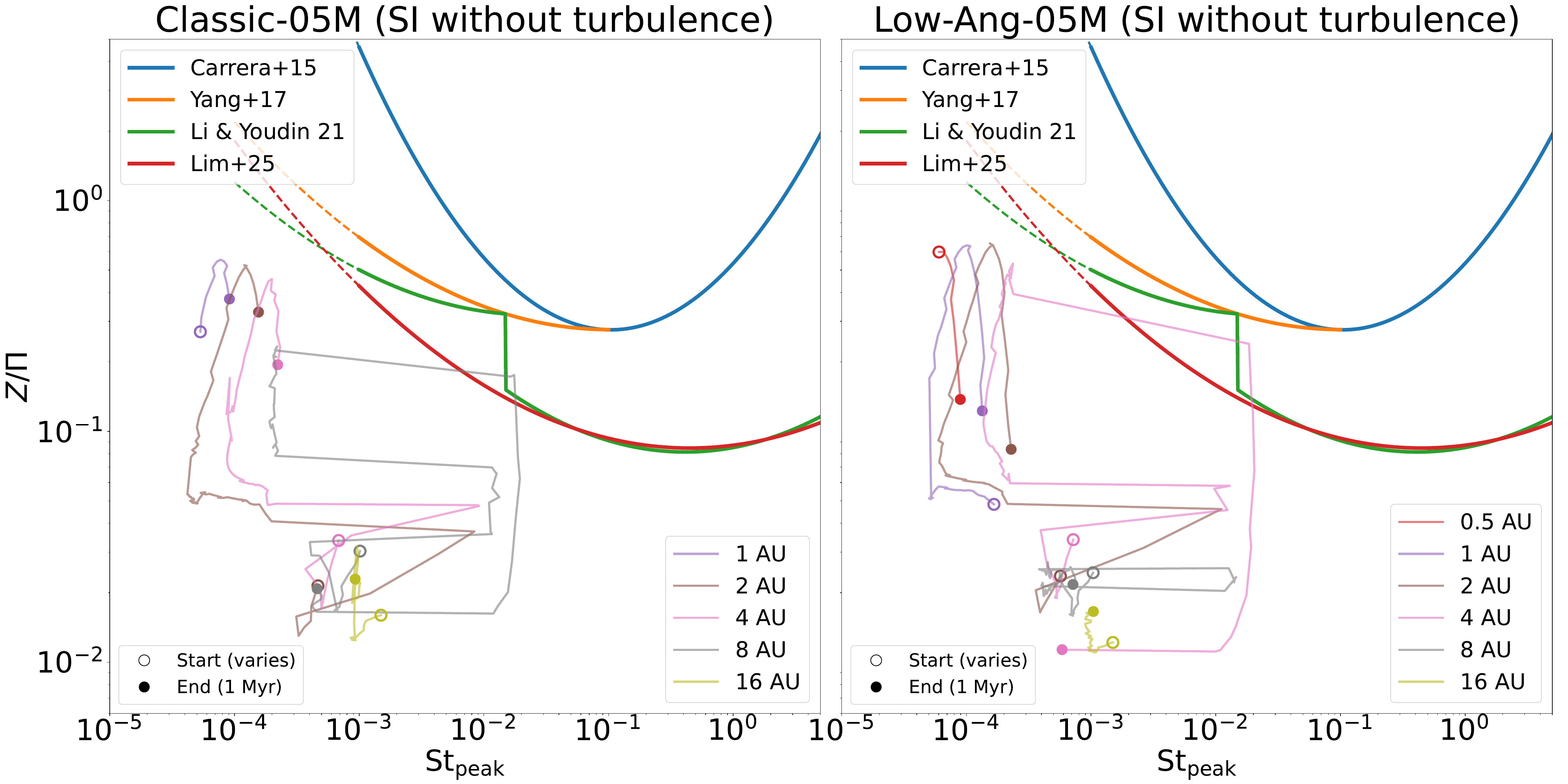}
\caption{Same as Fig.~\ref{fig:eps_vs_St_05M} but showing the SI criteria that do not account for external turbulence \citep{Lim_2025,Li_2021,Yang_2017,Carrera_2015}. Note the use of $Z/\Pi$ on the vertical axis in this figure instead of $\epsilon$. Here $\Pi$ is a ``headwind parameter'' that drives SI self-turbulence \citep[for details, see those papers and][]{Sekiya_2018}.\label{fig:Z_PI_vs_St}}
\end{figure*}

At each radial location in the disk we computed the mean Stokes number, weighted by mass, $\St_{\rm mean}$, and the Stokes number of the particle mass bin that contains the highest total mass, $\St_{\rm peak}$.

\begin{eqnarray}
    \St_{\rm mean} &=&
        \frac{\sum_i \St_i \sigma_i}{\sum_i \sigma_i}\\
    \St_{\rm peak} &=&
        \St_i \;\;\;\;\;\;\text{where}\;\;\;\;\;\;i = {\rm argmax}(\sigma_{\rm i})\label{eqn:St_peak}
\end{eqnarray}

\noindent
where $\St_i$ is the Stokes number of particles in mass bin $i$, and $\sigma_{\rm i}$ is the mass density in that mass bin.

Figure \ref{fig:SPD_St} shows spacetime diagrams of $\St_{\rm peak}$ for the \texttt{Classic-M05} and \texttt{Low-Ang-M05} models. We omit the plots for $\St_{\rm mean}$ because they look nearly the same as those of $\St_{\rm peak}$. We also omit the \texttt{Soft-Land-M05} model as it is nearly the same as \texttt{Low-Ang-M05}. Importantly, we also remove the region interior to the silicate snowline as all silicates are in the form of vapor (strictly speaking there are CAIs, but they have a negligible contribution to the metal content of the disk).

The most prominent feature of Figure \ref{fig:SPD_St} is the two sharp changes in particle size. These correspond to the water snowline and the colder point just outside of the snowline where water ice becomes less sticky \citep{Musiolik_2019}. This creates a relatively narrow band with larger grains.

Figure \ref{fig:eps_vs_St_05M} shows the evolution of the midplane dust-to-gas ratio $\epsilon$ and the characteristic grain size $\St_{\rm peak}$ at various locations in the disk. The red curve is the SI criterion of \citet{Lim_2024} with $\alphavis = 10^{-3}$.\footnote{In \citet{Lim_2024}, the turbulence parameter was simply $\alpha$. However, it is equivalent to our $\alphavis$} This is the first study of where in particle size vs. dust-to-gas ratio parameter space SI filaments reach the Roche density in the presence of external turbulence. Here we use the criterion for $\alphavis = 10^{-3}$, in line with our simulations. The dashed red line is an extrapolation of the SI criterion beyond the range of the parameters examined by \citet{Lim_2024}.\footnote{While this is indeed an extrapolation and should be taken with caution, it is clear from the Figure that particle sizes do not even come close to the planetesimal-forming region}

Next, the purple dotted line in Figure \ref{fig:eps_vs_St_05M} is the $\epsilon = 0.3$ threshold obtained by \citet{Carrera_2025}, who derived a semi-analytic expression to account for the interaction between the SI and dust growth. The key idea behind that work is that SI filaments dampen turbulence due to particle feedback; this promotes grain growth, which in turn makes the SI more effective. They found that, as long as the midplane dust-to-gas ratio can reach $\epsilon = 0.3$, this \textit{concentration-coagulation} feedback loop can bring SI filaments into the planetesimal formation region of \citet{Lim_2024}. Figure \ref{fig:eps_vs_St_05M} shows that the $(\epsilon,\St)$ in the disk are at least an order of magnitude away from what is needed for this concentration-coagulation mechanism to work.

As a side-note, several particle tracks in Figure \ref{fig:eps_vs_St_05M} show a sharp turn to the right (i.e., toward higher $\St$) followed by a decline in $\epsilon$. This happens when the water snowline passes through that location in the disk.

Figure \ref{fig:Z_PI_vs_St} is the same as Figure \ref{fig:eps_vs_St_05M}, but instead showing the various SI criteria that do not account for the effect of turbulence on the SI. Comparing these figures shows two things:

\begin{itemize}
\item Our results show that, under realistic conditions for a Class 0/I disk, even taking into account both localized dust enhancements at the water snowline or the outer edge of an inflationary disk, as well as the concentration-coagulation feedback loop of \citet{Carrera_2025}, planetesimal formation via the SI is far out of reach. Even a moderate amount of turbulence ($\alphavis = 10^{-3}$) has a significant deleterious effect on the SI. Turbulence decreases the midplane dust concentration and \citet{Lim_2024} showed that it also interferes with the SI's ability to form dense particle filament.

\item Even the SI criteria that do not account for external turbulence at all \citep{Carrera_2015,Yang_2017,Li_2021,Lim_2025} are mostly inconsistent with planetesimal formation in any but a handful of locations.
\end{itemize}

\subsection{Effect of Initial Stellar Mass}
\label{sec:results:stellar_mass}

\citet{Morbidelli_2022} and \citet{Marschall_2023} conducted simulations with $M_\star = 0.5\Msun$, so it is important that our first set of simulations followed that example to facilitate comparison. However, as these are very young systems, it is worthwhile exploring smaller initial stellar masses. For example, the stellar mass determines the centrifugal radius, the gravitational potential energy of infall, and the region of the disk where gas or dust flow outward.

\begin{figure*}[!ht]
\centering
\includegraphics[width=1\textwidth]{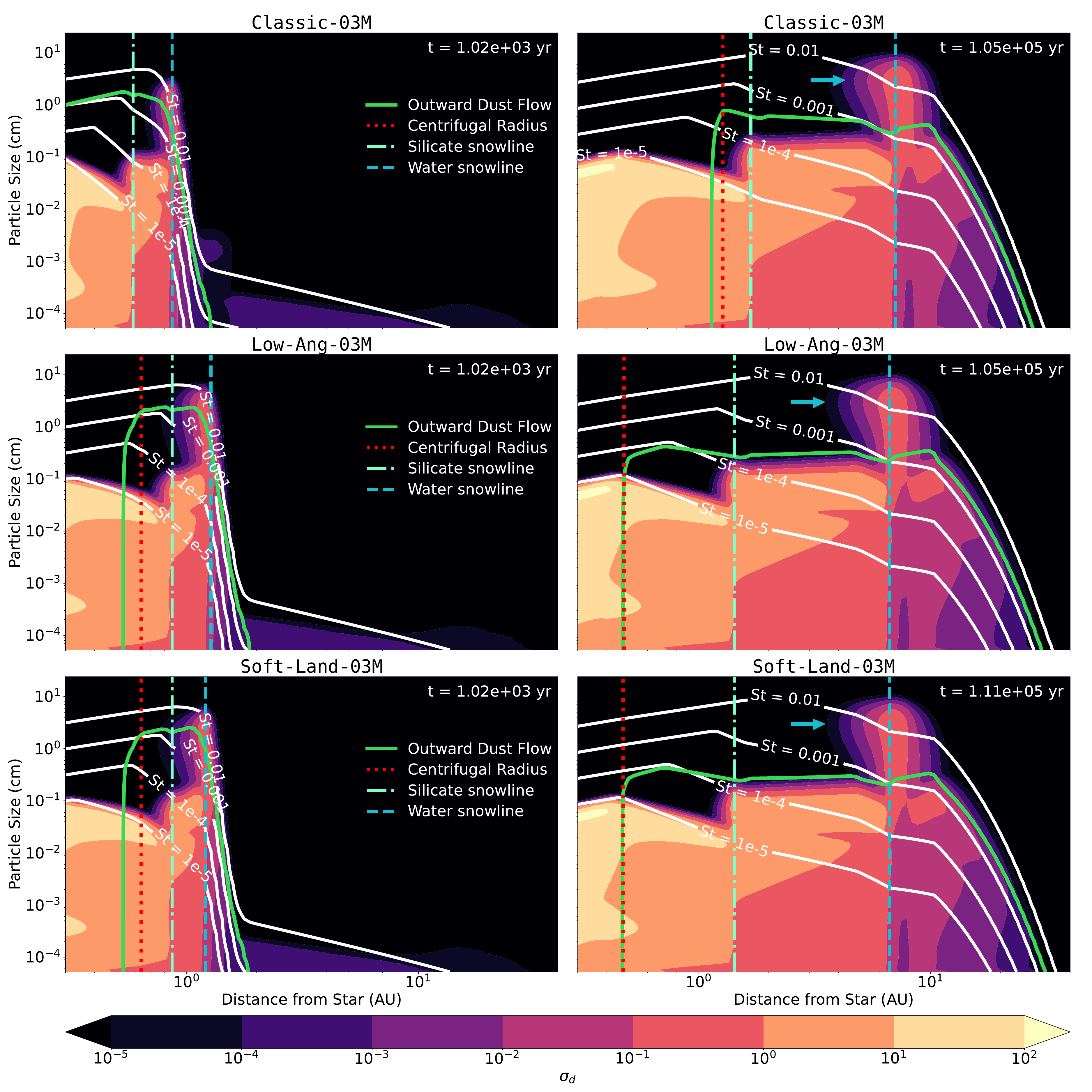}
\caption{Same as Fig.~\ref{fig:dust_density} but with the $M_{\star,0} = 0.3\Msun$ models. The results are similar to the ones with $M_{\star,0} = 0.5\Msun$. The stellar mass affects the centrifugal radius, which in turn affects the size of the disk and the radial extent of the region with outward dust flow. The \texttt{Classic-M03} model is much smaller and colder than \texttt{Classic-M05}, while the others become slightly warmer. Aside from that, the changes are generally minor and do not seem to fundamentally change any of the key results.\label{fig:dust_density_03M}}
\end{figure*}

\begin{figure*}[h]
\centering
\includegraphics[width=1\textwidth]{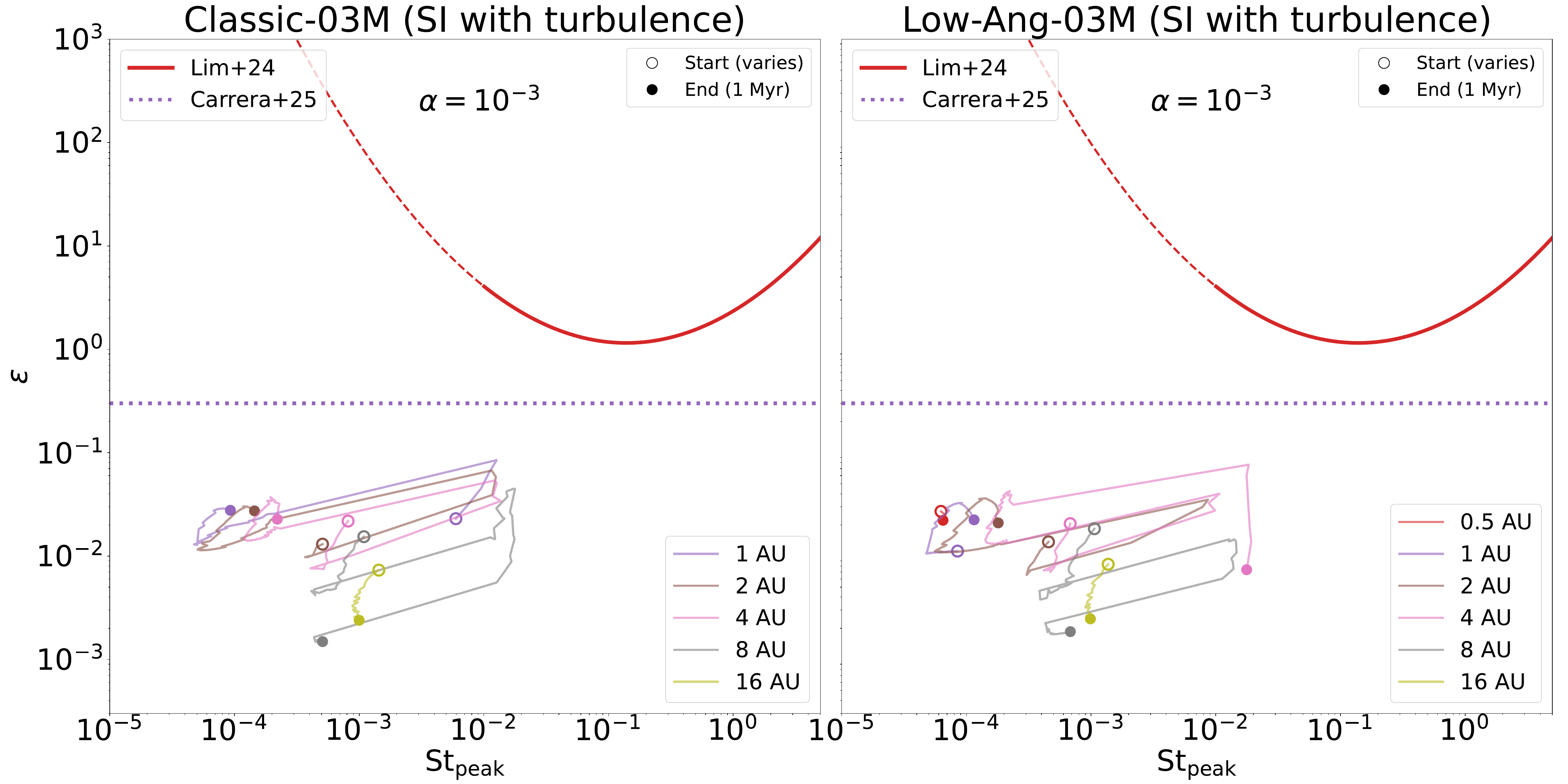}
\caption{Same as Fig.~\ref{fig:eps_vs_St_05M} but for the \texttt{Classic-M03} and \texttt{Low-Ang-M03} models.\label{fig:eps_vs_St_03M}}
\end{figure*}

Figure \ref{fig:dust_density_03M} shows snapshots of the surface density of grains for the runs with an initial stellar mass of 0.3$\Msun$, taken at the same times as those in Figure \ref{fig:dust_density}. The figures look quite similar. The biggest differences are seen in the early snapshots ($t = 10^3\yr$) --- most notably, the \texttt{Classic} model starts out smaller and colder --- but by $t = 10^5\yr$ it is somewhat difficult to tell the two sets of simulations apart. There are some differences in the radial extent of the disk and the extent of the region with outward dust flow, but they do not affect the key results in any significant way. The spacetime diagrams all look very similar. In Figure \ref{fig:eps_vs_St_03M} we see that the particle tracks are quite similar to those in Figure \ref{fig:eps_vs_St_05M}, including the main result that the conditions in the disk are at least an order of magnitude away from what is required to form planetesimals by the SI in a turbulent medium, even considering the criteria for the \textit{concentration-coagulation} feedback loop proposed by \citet{Carrera_2025}.

%
%
\section{Discussion}
\label{sec:discussion}

\subsection{Comparison with Previous Works}
\label{sec:discussion:estrada}

\citet{Estrada_2023} conducted an investigation of dust growth in a disk without infall and reached similar conclusions to ours. In particular, their results for $\alpha = 10^{-3}$  are very similar to our particle trajectories for the SI criteria without external turbulence (Figure \ref{fig:Z_PI_vs_St}).\footnote{For simplicity of notation, we dropped the ``vis" subscript on $\alpha$ in this section} One important contribution of their work is that they also explored $\alpha = 10^{-4}$ and found that the larger grains still did not reach the SI criterion. In comparison, our work adds infall and we focus on the more recent works on how the SI criterion is affected by external turbulence.

Additionally, upon finishing this paper, a recent paper came out on the effect of MHD disk winds on the early evolution of Class 0/I disks and the growth of dust grains \citep{Kawasaki_2025}. We discuss this paper more in Section~\ref{sec:discussion:winds} when we talk about winds. However, here we point out that while the exact details of their models differ from ours, \citet{Kawasaki_2025} find that assuming both magnetically driven turbulence via the magnetorotational instability (MRI; \citealt{Balbus_1998}) and gravitoturbulence (which played a larger role in their simulations due to an increased radial extent of their disk) driving accretion, planetesimal formation is also significantly limited.

Our model and those in \citet{Estrada_2023}, \citet{Morbidelli_2022}, and \citet{Kawasaki_2025} all differ in some way. 

\begin{itemize}
    \item Like \citet{Estrada_2023} and \citet{Kawasaki_2025}, we have a dust evolution model that tracks the entire particle size distribution.
    \item Like \citet{Morbidelli_2022} and \citet{Kawasaki_2025}, we model a Class 0/I inflationary disk with infall.
    \item Unlike our work, all three of these works track vapor species and condensation.
\end{itemize}

Thus, while we do not fully understand why our results are discrepant with \citet{Morbidelli_2022}, our hypothesis is that solving for the entire size distribution of particles (as is done with \texttt{DustPy} here) is the critical ingredient. However, more work is needed to test this notion, which we leave open for a future investigation.

\subsection{Open Questions Related to the alpha Model}
\label{sec:discussion:alpha_model}
In their seminal work, \citet{Shakura_1973} advanced a physical argument that angular momentum transport caused by magnetic and turbulent stress is proportional to the gas pressure.  In this so-called ``$\alpha$~model", disk viscosity is given by

\begin{equation}
    \nu = \alpha \frac{\cs^2}{\Omega}
        \propto \alpha \frac{T}{\Omega}
\end{equation}

\noindent
It is also common to choose constant $\alpha$, which then makes $\nu$ linear with $T$. This is certainly true for our work (except for a negligible contribution from self-gravity) and related works such as \citet{Morbidelli_2022,Marschall_2023,Estrada_2023}. Because so many 1D models rely on the $\alpha$ model's assumption that viscosity is dependent on temperature, it is worth exploring how this assumption might affect our results. For example, a linear dependence $T$ and $\nu$ has not been robustly established. Focusing on the magnetorotational instability (MRI; \citealt{Balbus_1998}), \citet{Sano_2004} and \citet{Lyra_2008} both found that the turbulent stress due to the MRI has a power-law dependence on pressure ($\propto P^n$) with an exponent of $n = 0.25$, instead of a linear one. But \citet{Minoshima_2015} argued that the exact relation depends on the numerical scheme, and \citet{Ross_2016} argued that the small $n$ is an artifact of small simulation domains. Instead, they found that the exponent depends on the magnetic flux and diffusivity, with a range from $n = 0.5$ to 0.9 (close to the $\alpha$ model). \citet{Shadmehri_2018} argued that $n < 0.5$ leads to unrealistically steep radial profiles for $\Sigma$ and $T$. All this is to say, the exact relationship between $\nu$ and $T$ in MRI-driven turbulence is an area of active research, and if it is very different from linear, that could impact our results. 

There are other purely hydrodynamic mechanisms that may also contribute to an effective $\alpha$, namely the Zombie Vortex Instability (ZVI; \citealt{Marcus_2015}), the Convective Overstability (COV; \citealt{Klahr_2014,Lyra_2014}), and the Vertical Shear Instability (VSI; \citealt{Nelson_2013}).\footnote{We don't include a discussion of gravitoturbulence within the context of the $\alpha$ model here since it plays a negligible role in our calculations.}. 

While these hydrodynamic mechanisms are intimately linked to the thermal properties of the gas (e.g., the cooling timescale controls which mechanism dominates), to our knowledge, the only detailed study of the temperature dependence of turbulent stresses from these hydrodynamic mechanisms is a focus on the VSI by \citet{Manger_2021}. They found that the stress induced by the VSI scales not with the pressure but instead with the square of the exponent of the radial temperature profile (for a limited range of exponent values). Thus, as with the MRI studies, the relationship between $\nu$ and $T$ from purely hydrodynamic mechanisms remains somewhat unclear.

\subsection{Disk Winds}
\label{sec:discussion:winds}

Perhaps the most important missing ingredient in our model is disk winds. Starting with the concept of $\alpha$ viscosity from \citet{Shakura_1973}, one can write $\alpha$ as the density-weighted Maxwell and Reynolds $r\phi$ component to the stress tensor \citep{Balbus_1998}

\begin{eqnarray}
    W_{r\phi} &\equiv& \frac{
        \langle \rho v_r \delta v_\phi - B_r B_\phi\rangle
    }{
        \langle \rho \cs^2 \rangle
    }\\
    \alpha &\equiv& \overline{|W_{r\phi}|}
\end{eqnarray}

\noindent
where $\phi$ is the azimuthal direction, $\delta v_\phi$ is the azimuthal velocity with Keplerian shear subtracted, and $\mathbf{B}$ is the magnetic field. In a steady-state disk where $W_{r\phi}$ is the only source of accretion, the accretion rate is given by

\begin{equation}
    \dot{M}_{R\phi} = \frac{2\pi\cs^2\Sigma}{\Omega}\alpha
\end{equation}

\noindent
Conversely, the accretion rate for a steady-state disk where the only source of accretion is a disk wind is defined via the $z \phi$ component of the stress tensor, $W_{z\phi}$ (see \citealt{Simon_2013}) as

\begin{eqnarray}\label{eqn:Mdot_zphi}
    \dot{M}_{z\phi} &=&
        \frac{8r\cs^2\Sigma}{H\Omega}
        \overline{|W_{z\phi}|}
        \sqrt{\frac{\pi}{2}}\\
    W_{z\phi} &\equiv&
        \left.
        \frac{\rho v_z \delta v_\phi - B_z B_\phi}{\rho_0\cs^2}
        \right|^{z_{\rm bw}}_{-z_{\rm bw}}
\end{eqnarray}

\noindent
where $\pm z_{\rm bw}$ are integration limits corresponding to the base of the wind on either side of the disk \citep[see][for details]{Simon_2013}. Note that Equation \ref{eqn:Mdot_zphi} differs from the expression in \citet{Simon_2013} by a factor of $\sqrt{2}$ because they define $H$ differently than we do.

With these ingredients, an approximate expression for the accretion rate in a disk with both viscous stress and disk winds is

\begin{equation}\label{eqn:M_dot_wind}
    \dot{M} = \frac{2\pi\cs^2\Sigma}{\Omega}
        \left[
            \alpha + \sqrt{\frac{8}{\pi}} \frac{r}{H}
            \overline{|W_{z\phi}|}
        \right].
\end{equation}

\noindent
Finally, \citet{Simon_2013} estimate that $|W_{z\phi}|$ is in the same magnitude range as $\alpha$. If so, given that $r/H \gg 1$, disk winds may be the dominant driver of accretion, even in the Class 0/I stage. There are a few ways that our results might change under wind-dominated accretion:

\begin{enumerate}
\item \textbf{Less viscous spreading.}

Winds contribute to accretion, but not to viscous spreading. A wind-dominated disk would spread less, which would reduce the amount of solid material advected across snowlines. This may pose an additional challenge for the SI. That said, even a wind-dominated disk could have significant spreading in the early GI-dominated phase. If so, the SI might be active only in these early stages.

\item \textbf{Less heating.}

By the same token, lower viscosity for a given $\dot{M}$ would make the disk colder. In our models viscous heating is the dominant source of heat in the inner disk. Removing a significant fraction of that would bring the silicate and water snowlines much closer to the star. However, $\St_{\rm frag}$ at the snowline would remain the same unless $\alpha$ is different (see the next point). Therefore, the implications for the SI are unclear.

\item \textbf{Lower $\alpha$?}

\begin{figure}[h]
\centering
\includegraphics[width=0.49\textwidth]{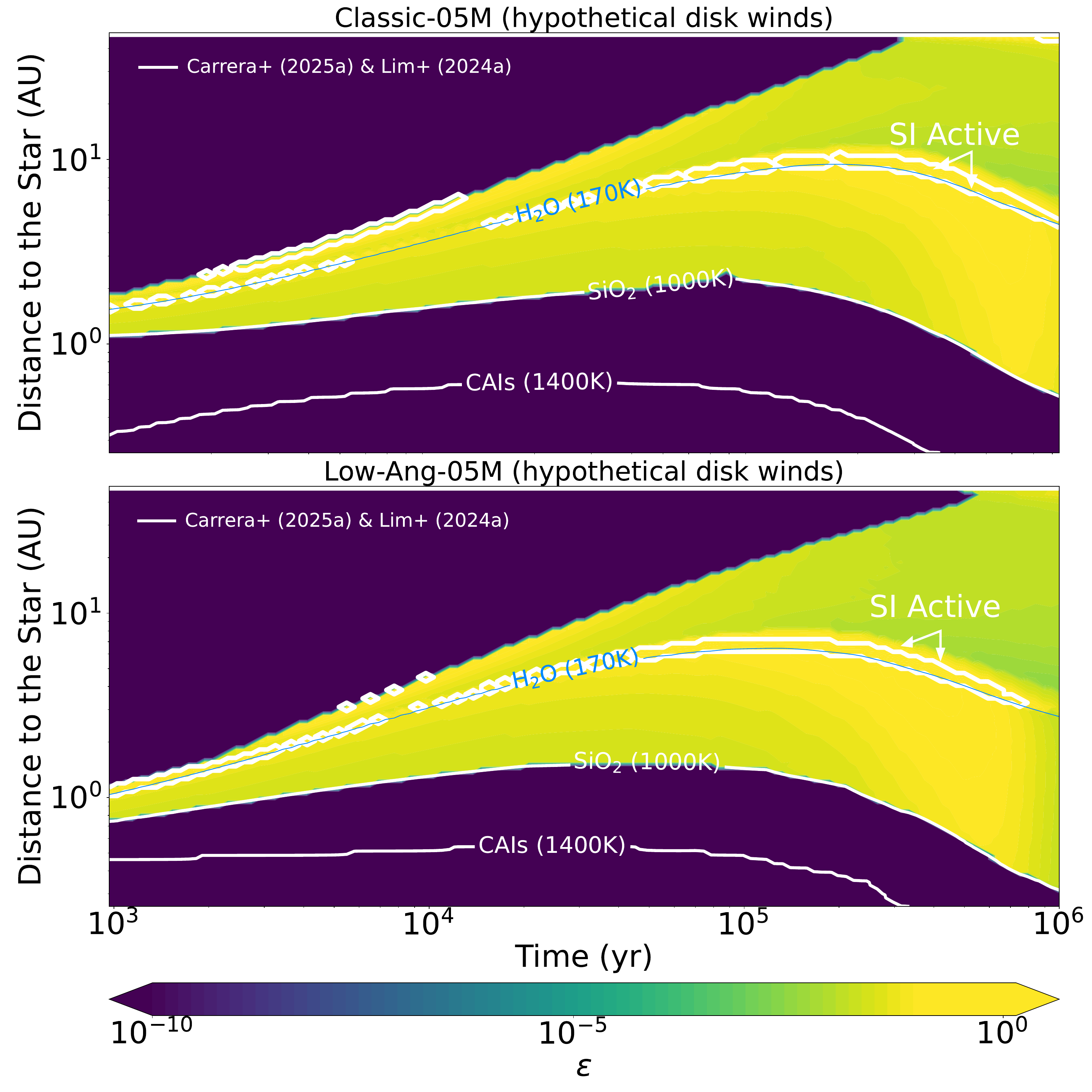}
\caption{Toy model where we increase $\St$ and $\epsilon$ by $\times 10$ (equivalent to decreasing $\alphavis$ to $10^{-4}$) but otherwise keep the disk structure intact. There are small regions, mainly around the water snowline and at the outer edge of the disk (though only for the first $\sim 10^4$ yr at the outer edge), where this model reaches the $\epsilon = 0.3$ threshold of \citet{Carrera_2025}. That is the threshold where their ``concentration-coagulation'' feedback loop can reach the SI criterion in the presence of turbulence.\label{fig:disk_wind_toy_model}}
\end{figure}

For any given disk lifetime or accretion rate $\dot{M}$, a disk with winds ($W_{z\phi}$) must have a lower $\alpha$. Note that disk lifetimes and accretion rates are not fully constrained, but a lower $\alpha$ would result in both larger $\St$ and greater midplane concentration $\epsilon$.

Since $\St_{\rm frag} \propto 1/\alpha$ and $\epsilon \propto \sqrt{\St/\alpha} \propto 1/\alpha$, all other things being equal, reducing $\alphavis$ to $10^{-4}$ would increase both $(\St,\epsilon)$ by a factor of 10. As a proof of concept, we can consider a toy model (Figure \ref{fig:disk_wind_toy_model}) where we do just that, but otherwise leave our disk structure intact. 

We find that in a wind-dominated disk, planetesimals may form around the water snowline and at the outer edge of the disk (though here, this only lasts for the first $\sim 10^4$ yr). However, the formation region does appear to be limited to these locations.

A couple of caveats with this toy model: Disk winds will surely alter the disk structure, and turbulence is not the only source of grain collisions --- for large grains, drift-induced velocities can dominate the collision speed (see the \texttt{DustPy} documentation).

However, despite its limitations, this thought experiment can give us some sense of how much disk winds might alter our results.
\end{enumerate}

As mentioned previously, upon writing of this manuscript, \citet{Kawasaki_2025} published a study of dust growth in a Class 0/I disk both with and without MHD winds. Indeed, they found that planetesimal formation via the SI should be significantly easier {\it with} an MHD wind than without. However, their equivalent of our Figs.~\ref{fig:Z_PI_vs_St} and \ref{fig:eps_vs_St_03M} (figures 16 and 17 in their paper) did not include the stricter SI criterion in the presence of turbulence (found by \citealt{Lim_2024}) even with an MHD wind. While MHD winds may very well dominate angular momentum transport, recent studies (e.g., \citealt{Cui_2021,Rea_2024}) have shown that turbulence can persist even when winds play an important role.

While we intend to include winds in a follow up study to investigate this further, these considerations do suggest that planetesimal formation via the SI may still be difficult in Class 0/I disks even if MHD winds dominate the evolution.

\subsection{Snowlines}
\label{sec:discussion:snowlines}

\texttt{DustPy} only allows a fairly simplified treatment of the water and silicate snowlines. In a realistic disk, solid material that drifts inward across one of the snowlines will evaporate and join the gas component. Since the gas accretion rate is much slower than solid drift, the evaporated material cannot be removed as quickly in its vapor form as it is supplied in solid form \citep[e.g.,][]{Cuzzi_2004}. This is not something that one can currently model with \texttt{DustPy}. However, we can speculate on how this concentration of vapor might affect our results:

\begin{itemize}
\item One possibility \citep[seen in][]{Morbidelli_2022}, is that vapor could spread back through the snowline. This can increase the total solid mass near the snowline, but it does not necessarily alter grain sizes.

\item A second possibility is that vapor condensation might produce larger grains beyond the snowline \citep[e.g.,][]{Ros_2019}. However, it is unlikely that the grains would be much larger than the ones in our models because the grain sizes in those models are fragmentation-limited \citep{Stammler_2017}.

\item Importantly, \citet{Estrada_2023} do track multi-species vapor and condensation, and their results resemble ours.

\item Finally, the accumulation of vapor inside the snowline leads to an increase in mean molecular weight, which lowers the sounds speed, which leads to the formation of a pressure bump \citep[e.g.,][]{Charnoz_2021,Drazkowska_2018,Drazkowska_2017}. However, this might not necessarily allow the SI to form planetesimals. \citet{Carrera_2022} conducted a large high-res simulation of the SI with a pressure bump with a particle Stokes number comparable to the ones in these simulations. They found that there was no indication of the SI. However, a sufficiently large pressure bump will form a particle trap and could form planetesimals by gravitational instability (i.e., if enough particles pile up, the local density will surpass the Roche density.)
\end{itemize}

\subsection{Accretion Model}
\label{sec:discussion:accretion_model}

As noted earlier in this paper, the simulations of \cite{Mauxion_2024} show a more complex infall geometry than what we use in our model. In particular, the simulations by \cite{Mauxion_2024} show that some of the infalling gas occurs along the mid-plane of the disk and shocks the edge of the disk, whereas the remaining infalling gas gently lands on the disk surface. This is at odds with 1D models like ours and those of \citet{Morbidelli_2022} and \citet{Marschall_2023} that insert material at the centrifugal radius without considering this more complex geometry. If infall occurs as it does in those simulations, then at least some of the gas is deposited wherever it encounters the edge of the disk, which will influence where shock heating occurs.

In addition, the simulations of \citet{Mauxion_2024} show a great deal of structure that is not captured by our current models and may be important for trapping particles. Furthermore, other MHD simulations (even without infall) also show disk substructures (e.g., zonal flows; \citealt{Bethune_2017,Hu_2019}). We leave the investigation of substructures in Class 0/I disks for future investigations.

%
%
\section{Summary and Conclusions}
\label{sec:conclusions}

In this work, we studied the growth of small dust grains into larger particles in a viscously expanding ``inflationary" disk and whether or not conditions can be met for the SI to work. Similar to other works, we find that the outward flow of gas counteracts the tendency of solid grains to drift toward the star, resulting in higher concentrations and larger grains. This is most evident at the water snowline, where an ``advection-condensation-drift'' loop leads to a significant solid pile up. However, we do not see a similar process at the silicate condensation line. Instead, we find that silicate grains just beyond the silicate condensation front are advected outward along with the gas. We do, however, find an additional solid pile up at the edge of the expanding disk that appears to not have been previously reported.

Our investigation stands out from other works in that we explore the implications of recent results on how the SI responds to turbulence. Specifically, \citet{Lim_2024} conducted 3D simulations of the SI in the presence of turbulence and \citet{Carrera_2025} derived analytic expressions for how the SI interacts with dust coagulation in a turbulent medium. Applying these developments to our disk model, we find that for modest $\alphavis = 10^{-3}$ turbulence, the planetesimal formation region is completely out of reach. The midplane dust-to-gas ratio is at least an order of magnitude too small to form planetesimals, even if the SI is assisted by its feedback loop with dust coagulation. Furthermore, the maximum grain size in our simulations remains quite small due to turbulent fragmentation.

All in all, the quest for a formation scenario in which realistic disk conditions are able to satisfy the conditions for the SI \citep{Carrera_2015,Yang_2017,Li_2021,Lim_2025}, continues. We still believe that Class 0/I disks and infall deserve careful attention, given the observational evidence that planets form early. An inflationary disk's ability to overcome the radial drift barrier is promising, but planetesimal formation via the SI appears to be thwarted by disk turbulence, which simultaneously keeps solid grains small and reduces the midplane dust-to-gas ratio. Future works should consider variations of this formation scenario that ideally have less turbulence, or perhaps exploring alternatives to the SI.

As discussed above, one promising route towards early planetesimal formation is magnetic wind-driven accretion, which (for a given accretion rate) implies a small $\alphavis$. A simple toy model where we allow grain sizes to grow due to reduced turbulent fragmentation does imply that planetesimal formation is possible in such a disk. However, the regions where this occurs are limited, and there are many simplifications of our toy model that may not hold true when a more self-consistent model is used.

Clearly, planetesimal formation occurs at some point. However, when and how this happens remains debated. Our results imply that planetesimal formation (at least via the SI) may be difficult unless some mechanism keeps turbulence low, and even then, the SI might be limited to the water snowline or other disk structures.


\section*{Acknowledgments}
\nolinenumbers

The authors thank Eric Gaidos, Wenrui Xu, Geoffroy Lesur, Tom Megeath, Kaitlin Kratter, and Lee Hartmann for their insightful comments and questions that greatly improved the quality of this work. DC and JBS acknowledge support from NASA under {\em Emerging Worlds} through grant 80NSSC21K0037, and {\em Exoplanets Research} through grants 80NSSC24K0959 and 80NSSC22K0267. CH acknowledges support from the National Science Foundation through NSF AAG grant No.8612407679. 

%
%

\bibliography{references}{}

\begin{thebibliography}{}
\expandafter\ifx\csname natexlab\endcsname\relax\def\natexlab#1{#1}\fi
\providecommand{\url}[1]{\href{#1}{#1}}
\providecommand{\dodoi}[1]{doi:~\href{http://doi.org/#1}{\nolinkurl{#1}}}
\providecommand{\doeprint}[1]{\href{http://ascl.net/#1}{\nolinkurl{http://ascl.net/#1}}}
\providecommand{\doarXiv}[1]{\href{https://arxiv.org/abs/#1}{\nolinkurl{https://arxiv.org/abs/#1}}}

\bibitem[{{Balbus} \& {Hawley}(1998)}]{Balbus_1998}
{Balbus}, S.~A., \& {Hawley}, J.~F. 1998, Reviews of Modern Physics, 70, 1,
  \dodoi{10.1103/RevModPhys.70.1}

\bibitem[{{Bell} \& {Lin}(1994)}]{Bell_1994}
{Bell}, K.~R., \& {Lin}, D.~N.~C. 1994, \apj, 427, 987, \dodoi{10.1086/174206}

\bibitem[{{B{\'e}thune} {et~al.}(2021){B{\'e}thune}, {Latter}, \&
  {Kley}}]{Bethune_2021}
{B{\'e}thune}, W., {Latter}, H., \& {Kley}, W. 2021, \aap, 650, A49,
  \dodoi{10.1051/0004-6361/202040094}

\bibitem[{{B{\'e}thune} {et~al.}(2017){B{\'e}thune}, {Lesur}, \&
  {Ferreira}}]{Bethune_2017}
{B{\'e}thune}, W., {Lesur}, G., \& {Ferreira}, J. 2017, Astronomy \&
  Astrophysics, 600, A75, \dodoi{10.1051/0004-6361/201630056}

\bibitem[{{Birnstiel} {et~al.}(2010){Birnstiel}, {Dullemond}, \&
  {Brauer}}]{Birnstiel_2010}
{Birnstiel}, T., {Dullemond}, C.~P., \& {Brauer}, F. 2010, \aap, 513, A79,
  \dodoi{10.1051/0004-6361/200913731}

\bibitem[{{Birnstiel} {et~al.}(2012){Birnstiel}, {Klahr}, \&
  {Ercolano}}]{Birnstiel_2012}
{Birnstiel}, T., {Klahr}, H., \& {Ercolano}, B. 2012, \aap, 539, A148,
  \dodoi{10.1051/0004-6361/201118136}

\bibitem[{{Bitsch} {et~al.}(2013){Bitsch}, {Crida}, {Morbidelli}, {Kley}, \&
  {Dobbs-Dixon}}]{Bitsch_2013}
{Bitsch}, B., {Crida}, A., {Morbidelli}, A., {Kley}, W., \& {Dobbs-Dixon}, I.
  2013, \aap, 549, A124, \dodoi{10.1051/0004-6361/201220159}

\bibitem[{{Blum} \& {Wurm}(2008)}]{Blum_2008}
{Blum}, J., \& {Wurm}, G. 2008, \araa, 46, 21,
  \dodoi{10.1146/annurev.astro.46.060407.145152}

\bibitem[{{Carrera} {et~al.}(2017){Carrera}, {Gorti}, {Johansen}, \&
  {Davies}}]{Carrera_2017}
{Carrera}, D., {Gorti}, U., {Johansen}, A., \& {Davies}, M.~B. 2017, \apj, 839,
  16, \dodoi{10.3847/1538-4357/aa6932}

\bibitem[{{Carrera} {et~al.}(2015){Carrera}, {Johansen}, \&
  {Davies}}]{Carrera_2015}
{Carrera}, D., {Johansen}, A., \& {Davies}, M.~B. 2015, \aap, 579, A43,
  \dodoi{10.1051/0004-6361/201425120}

\bibitem[{{Carrera} {et~al.}(2025){Carrera}, {Lim}, {Eriksson}, {Lyra}, \&
  {Simon}}]{Carrera_2025}
{Carrera}, D., {Lim}, J., {Eriksson}, L. E.~J., {Lyra}, W., \& {Simon}, J.~B.
  2025, arXiv e-prints, arXiv:2503.03105, \dodoi{10.48550/arXiv.2503.03105}

\bibitem[{{Carrera} \& {Simon}(2022)}]{Carrera_2022}
{Carrera}, D., \& {Simon}, J.~B. 2022, \apjl, 933, L10,
  \dodoi{10.3847/2041-8213/ac6b3e}

\bibitem[{{Carrera} {et~al.}(2021){Carrera}, {Simon}, {Li}, {Kretke}, \&
  {Klahr}}]{Carrera_2021}
{Carrera}, D., {Simon}, J.~B., {Li}, R., {Kretke}, K.~A., \& {Klahr}, H. 2021,
  \aj, 161, 96, \dodoi{10.3847/1538-3881/abd4d9}

\bibitem[{{Chambers}(2001)}]{Chambers_2001}
{Chambers}, J.~E. 2001, \icarus, 152, 205, \dodoi{10.1006/icar.2001.6639}

\bibitem[{{Charnoz} {et~al.}(2021){Charnoz}, {Avice}, {Hyodo}, {Pignatale}, \&
  {Chaussidon}}]{Charnoz_2021}
{Charnoz}, S., {Avice}, G., {Hyodo}, R., {Pignatale}, F.~C., \& {Chaussidon},
  M. 2021, \aap, 652, A35, \dodoi{10.1051/0004-6361/202038797}

\bibitem[{{Connelly} {et~al.}(2012){Connelly}, {Bizzarro}, {Krot}, {Nordlund},
  {Wielandt}, \& {Ivanova}}]{Connelly_2012}
{Connelly}, J.~N., {Bizzarro}, M., {Krot}, A.~N., {et~al.} 2012, Science, 338,
  651, \dodoi{10.1126/science.1226919}

\bibitem[{{Cossins} {et~al.}(2010){Cossins}, {Lodato}, \&
  {Clarke}}]{Cossins_2010}
{Cossins}, P., {Lodato}, G., \& {Clarke}, C. 2010, \mnras, 401, 2587,
  \dodoi{10.1111/j.1365-2966.2009.15835.x}

\bibitem[{{Cossins} {et~al.}(2009){Cossins}, {Lodato}, \&
  {Clarke}}]{Cossins_2009}
{Cossins}, P., {Lodato}, G., \& {Clarke}, C.~J. 2009, \mnras, 393, 1157,
  \dodoi{10.1111/j.1365-2966.2008.14275.x}

\bibitem[{{Cui} \& {Bai}(2021)}]{Cui_2021}
{Cui}, C., \& {Bai}, X.-N. 2021, \mnras, 507, 1106,
  \dodoi{10.1093/mnras/stab2220}

\bibitem[{{Cuzzi} {et~al.}(2008){Cuzzi}, {Hogan}, \& {Shariff}}]{Cuzzi_2008}
{Cuzzi}, J.~N., {Hogan}, R.~C., \& {Shariff}, K. 2008, \apj, 687, 1432,
  \dodoi{10.1086/591239}

\bibitem[{{Cuzzi} \& {Zahnle}(2004)}]{Cuzzi_2004}
{Cuzzi}, J.~N., \& {Zahnle}, K.~J. 2004, \apj, 614, 490, \dodoi{10.1086/423611}

\bibitem[{{Desch} {et~al.}(2018){Desch}, {Kalyaan}, \& {O'D.
  Alexander}}]{Desch_2018}
{Desch}, S.~J., {Kalyaan}, A., \& {O'D. Alexander}, C.~M. 2018, \apjs, 238, 11,
  \dodoi{10.3847/1538-4365/aad95f}

\bibitem[{{Drazkowska} \& {Alibert}(2017)}]{Drazkowska_2017}
{Drazkowska}, J., \& {Alibert}, Y. 2017, \aap, 608, A92,
  \dodoi{10.1051/0004-6361/201731491}

\bibitem[{{Drazkowska} \& {Dullemond}(2018)}]{Drazkowska_2018}
{Drazkowska}, J., \& {Dullemond}, C.~P. 2018, \aap, 614, A62,
  \dodoi{10.1051/0004-6361/201732221}

\bibitem[{{Durisen} {et~al.}(2007){Durisen}, {Boss}, {Mayer}, {Nelson},
  {Quinn}, \& {Rice}}]{Durisen_2007}
{Durisen}, R.~H., {Boss}, A.~P., {Mayer}, L., {et~al.} 2007, in Protostars and
  Planets V, ed. B.~{Reipurth}, D.~{Jewitt}, \& K.~{Keil}, 607,
  \dodoi{10.48550/arXiv.astro-ph/0603179}

\bibitem[{{Estrada} \& {Umurhan}(2023)}]{Estrada_2023}
{Estrada}, P.~R., \& {Umurhan}, O.~M. 2023, \apj, 946, 15,
  \dodoi{10.3847/1538-4357/acb7db}

\bibitem[{{Fraser} {et~al.}(2017){Fraser}, {Bannister}, {Pike}, {Marsset},
  {Schwamb}, {Kavelaars}, {Lacerda}, {Nesvorn{\'y}}, {Volk}, {Delsanti},
  {Benecchi}, {Lehner}, {Noll}, {Gladman}, {Petit}, {Gwyn}, {Chen}, {Wang},
  {Alexandersen}, {Burdullis}, {Sheppard}, \& {Trujillo}}]{Fraser_2017}
{Fraser}, W.~C., {Bannister}, M.~T., {Pike}, R.~E., {et~al.} 2017, Nature
  Astronomy, 1, 0088, \dodoi{10.1038/s41550-017-0088}

\bibitem[{{Gammie}(2001)}]{Gammie_2001}
{Gammie}, C.~F. 2001, \apj, 553, 174, \dodoi{10.1086/320631}

\bibitem[{{Goldreich} \& {Lynden-Bell}(1965)}]{Goldreich_1965}
{Goldreich}, P., \& {Lynden-Bell}, D. 1965, \mnras, 130, 125,
  \dodoi{10.1093/mnras/130.2.125}

\bibitem[{{Gundlach} \& {Blum}(2015)}]{Gundlach_2015}
{Gundlach}, B., \& {Blum}, J. 2015, \apj, 798, 34,
  \dodoi{10.1088/0004-637X/798/1/34}

\bibitem[{{Hu} {et~al.}(2019){Hu}, {Zhu}, {Okuzumi}, {Bai}, {Wang}, {Tomida},
  \& {Stone}}]{Hu_2019}
{Hu}, X., {Zhu}, Z., {Okuzumi}, S., {et~al.} 2019, The Astrophysical Journal,
  885, 36, \dodoi{10.3847/1538-4357/ab44cb}

\bibitem[{{Johansen} {et~al.}(2007){Johansen}, {Oishi}, {Mac Low}, {Klahr},
  {Henning}, \& {Youdin}}]{Johansen_2007a}
{Johansen}, A., {Oishi}, J.~S., {Mac Low}, M.-M., {et~al.} 2007, \nat, 448,
  1022, \dodoi{10.1038/nature06086}

\bibitem[{{Johansen} \& {Youdin}(2007)}]{Johansen_2007b}
{Johansen}, A., \& {Youdin}, A. 2007, \apj, 662, 627, \dodoi{10.1086/516730}

\bibitem[{{Kawasaki} \& {Machida}(2025)}]{Kawasaki_2025}
{Kawasaki}, Y., \& {Machida}, M.~N. 2025, arXiv e-prints, arXiv:2503.19219,
  \dodoi{10.48550/arXiv.2503.19219}

\bibitem[{{Klahr} \& {Hubbard}(2014)}]{Klahr_2014}
{Klahr}, H., \& {Hubbard}, A. 2014, The Astrophysical Journal, 788, 21,
  \dodoi{10.1088/0004-637X/788/1/21}

\bibitem[{{Kokubo} \& {Ida}(1996)}]{Kokubo_1996}
{Kokubo}, E., \& {Ida}, S. 1996, \icarus, 123, 180,
  \dodoi{10.1006/icar.1996.0148}

\bibitem[{{Kokubo} \& {Ida}(2000)}]{Kokubo_2000}
---. 2000, \icarus, 143, 15, \dodoi{10.1006/icar.1999.6237}

\bibitem[{{Kratter} \& {Lodato}(2016)}]{Kratter_2016}
{Kratter}, K., \& {Lodato}, G. 2016, \araa, 54, 271,
  \dodoi{10.1146/annurev-astro-081915-023307}

\bibitem[{{Kruijer} {et~al.}(2017){Kruijer}, {Burkhardt}, {Budde}, \&
  {Kleine}}]{Kruijer_2017}
{Kruijer}, T.~S., {Burkhardt}, C., {Budde}, G., \& {Kleine}, T. 2017,
  Proceedings of the National Academy of Science, 114, 6712,
  \dodoi{10.1073/pnas.1704461114}

\bibitem[{{Lau} {et~al.}(2022){Lau}, {Drazkowska}, {Stammler}, {Birnstiel}, \&
  {Dullemond}}]{Lau_2022}
{Lau}, T. C.~H., {Drazkowska}, J., {Stammler}, S.~M., {Birnstiel}, T., \&
  {Dullemond}, C.~P. 2022, \aap, 668, A170, \dodoi{10.1051/0004-6361/202244864}

\bibitem[{{Laughlin} \& {Bodenheimer}(1994)}]{Laughlin_1994}
{Laughlin}, G., \& {Bodenheimer}, P. 1994, \apj, 436, 335,
  \dodoi{10.1086/174909}

\bibitem[{{Li} \& {Youdin}(2021)}]{Li_2021}
{Li}, R., \& {Youdin}, A.~N. 2021, \apj, 919, 107,
  \dodoi{10.3847/1538-4357/ac0e9f}

\bibitem[{{Lim} {et~al.}(2025){Lim}, {Simon}, {Li}, {Carrera}, {Baronett},
  {Youdin}, {Lyra}, \& {Yang}}]{Lim_2025}
{Lim}, J., {Simon}, J.~B., {Li}, R., {et~al.} 2025, \apj, 981, 160,
  \dodoi{10.3847/1538-4357/adb311}

\bibitem[{{Lim} {et~al.}(2024){Lim}, {Simon}, {Li}, {Armitage}, {Carrera},
  {Lyra}, {Rea}, {Yang}, \& {Youdin}}]{Lim_2024}
---. 2024, \apj, 969, 130, \dodoi{10.3847/1538-4357/ad47a2}

\bibitem[{{Lin} \& {Pringle}(1987)}]{Lin_1987}
{Lin}, D.~N.~C., \& {Pringle}, J.~E. 1987, \mnras, 225, 607,
  \dodoi{10.1093/mnras/225.3.607}

\bibitem[{{Liu} {et~al.}(2024){Liu}, {Roussel}, {Linz}, {Fang}, {Wolf},
  {Kirchschlager}, {Henning}, {Yang}, {Du}, {Flock}, \& {Wang}}]{Liu_2024}
{Liu}, Y., {Roussel}, H., {Linz}, H., {et~al.} 2024, arXiv e-prints,
  arXiv:2411.00277, \dodoi{10.48550/arXiv.2411.00277}

\bibitem[{{Lodato} \& {Rice}(2005)}]{Lodato_2005}
{Lodato}, G., \& {Rice}, W.~K.~M. 2005, \mnras, 358, 1489,
  \dodoi{10.1111/j.1365-2966.2005.08875.x}

\bibitem[{{Lynden-Bell} \& {Pringle}(1974)}]{Lynden_Bell_1974}
{Lynden-Bell}, D., \& {Pringle}, J.~E. 1974, \mnras, 168, 603,
  \dodoi{10.1093/mnras/168.3.603}

\bibitem[{{Lyra}(2014)}]{Lyra_2014}
{Lyra}, W. 2014, The Astrophysical Journal, 789, 77,
  \dodoi{10.1088/0004-637X/789/1/77}

\bibitem[{{Lyra} {et~al.}(2008){Lyra}, {Johansen}, {Klahr}, \&
  {Piskunov}}]{Lyra_2008}
{Lyra}, W., {Johansen}, A., {Klahr}, H., \& {Piskunov}, N. 2008, \aap, 479,
  883, \dodoi{10.1051/0004-6361:20077948}

\bibitem[{{Manara} {et~al.}(2018){Manara}, {Morbidelli}, \&
  {Guillot}}]{Manara_2018}
{Manara}, C.~F., {Morbidelli}, A., \& {Guillot}, T. 2018, \aap, 618, L3,
  \dodoi{10.1051/0004-6361/201834076}

\bibitem[{{Manger} {et~al.}(2021){Manger}, {Pfeil}, \& {Klahr}}]{Manger_2021}
{Manger}, N., {Pfeil}, T., \& {Klahr}, H. 2021, \mnras, 508, 5402,
  \dodoi{10.1093/mnras/stab2599}

\bibitem[{{Marcus} {et~al.}(2015){Marcus}, {Pei}, {Jiang}, {Barranco},
  {Hassanzadeh}, \& {Lecoanet}}]{Marcus_2015}
{Marcus}, P.~S., {Pei}, S., {Jiang}, C.-H., {et~al.} 2015, The Astrophysical
  Journal, 808, 87, \dodoi{10.1088/0004-637X/808/1/87}

\bibitem[{{Marschall} \& {Morbidelli}(2023)}]{Marschall_2023}
{Marschall}, R., \& {Morbidelli}, A. 2023, \aap, 677, A136,
  \dodoi{10.1051/0004-6361/202346616}

\bibitem[{{Mathis} {et~al.}(1977){Mathis}, {Rumpl}, \&
  {Nordsieck}}]{Mathis_1977}
{Mathis}, J.~S., {Rumpl}, W., \& {Nordsieck}, K.~H. 1977, \apj, 217, 425,
  \dodoi{10.1086/155591}

\bibitem[{{Mauxion} {et~al.}(2024){Mauxion}, {Lesur}, \&
  {Maret}}]{Mauxion_2024}
{Mauxion}, J., {Lesur}, G., \& {Maret}, S. 2024, \aap, 686, A253,
  \dodoi{10.1051/0004-6361/202348405}

\bibitem[{{Minoshima} {et~al.}(2015){Minoshima}, {Hirose}, \&
  {Sano}}]{Minoshima_2015}
{Minoshima}, T., {Hirose}, S., \& {Sano}, T. 2015, \apj, 808, 54,
  \dodoi{10.1088/0004-637X/808/1/54}

\bibitem[{{Morbidelli} {et~al.}(2022){Morbidelli}, {Bailli{\'e}}, {Batygin},
  {Charnoz}, {Guillot}, {Rubie}, \& {Kleine}}]{Morbidelli_2022}
{Morbidelli}, A., {Bailli{\'e}}, K., {Batygin}, K., {et~al.} 2022, Nature
  Astronomy, 6, 72, \dodoi{10.1038/s41550-021-01517-7}

\bibitem[{{Morbidelli} {et~al.}(2009){Morbidelli}, {Bottke}, {Nesvorn{\'y}}, \&
  {Levison}}]{Morbidelli_2009}
{Morbidelli}, A., {Bottke}, W.~F., {Nesvorn{\'y}}, D., \& {Levison}, H.~F.
  2009, \icarus, 204, 558, \dodoi{10.1016/j.icarus.2009.07.011}

\bibitem[{{Morbidelli} {et~al.}(2000){Morbidelli}, {Chambers}, {Lunine},
  {Petit}, {Robert}, {Valsecchi}, \& {Cyr}}]{Morbidelli_2000}
{Morbidelli}, A., {Chambers}, J., {Lunine}, J.~I., {et~al.} 2000, \maps, 35,
  1309, \dodoi{10.1111/j.1945-5100.2000.tb01518.x}

\bibitem[{{Mulders} {et~al.}(2021){Mulders}, {Pascucci}, {Ciesla}, \&
  {Fernandes}}]{Mulders_2021}
{Mulders}, G.~D., {Pascucci}, I., {Ciesla}, F.~J., \& {Fernandes}, R.~B. 2021,
  \apj, 920, 66, \dodoi{10.3847/1538-4357/ac178e}

\bibitem[{{Musiolik}(2021)}]{Musiolik_2021}
{Musiolik}, G. 2021, \mnras, 506, 5153, \dodoi{10.1093/mnras/stab1963}

\bibitem[{{Musiolik} \& {Wurm}(2019)}]{Musiolik_2019}
{Musiolik}, G., \& {Wurm}, G. 2019, \apj, 873, 58,
  \dodoi{10.3847/1538-4357/ab0428}

\bibitem[{{Nelson} {et~al.}(2013){Nelson}, {Gressel}, \&
  {Umurhan}}]{Nelson_2013}
{Nelson}, R.~P., {Gressel}, O., \& {Umurhan}, O.~M. 2013, Monthly Notices of
  the Royal Astronomical Society, 435, 2610, \dodoi{10.1093/mnras/stt1475}

\bibitem[{{Nesvorn{\'y}} {et~al.}(2021){Nesvorn{\'y}}, {Li}, {Simon}, {Youdin},
  {Richardson}, {Marschall}, \& {Grundy}}]{Nesvorny_2021}
{Nesvorn{\'y}}, D., {Li}, R., {Simon}, J.~B., {et~al.} 2021, The Planetary
  Science Journal, 2, 27, \dodoi{10.3847/PSJ/abd858}

\bibitem[{{Nesvorn{\'y}} {et~al.}(2019){Nesvorn{\'y}}, {Li}, {Youdin}, {Simon},
  \& {Grundy}}]{Nesvorny_2019}
{Nesvorn{\'y}}, D., {Li}, R., {Youdin}, A.~N., {Simon}, J.~B., \& {Grundy},
  W.~M. 2019, Nature Astronomy, 3, 808, \dodoi{10.1038/s41550-019-0806-z}

\bibitem[{{Nesvorn{\'y}} {et~al.}(2010){Nesvorn{\'y}}, {Youdin}, \&
  {Richardson}}]{Nesvorny_2010}
{Nesvorn{\'y}}, D., {Youdin}, A.~N., \& {Richardson}, D.~C. 2010, \aj, 140,
  785, \dodoi{10.1088/0004-6256/140/3/785}

\bibitem[{{{\"O}berg} \& {Wordsworth}(2019)}]{Oberg_2019}
{{\"O}berg}, K.~I., \& {Wordsworth}, R. 2019, \aj, 158, 194,
  \dodoi{10.3847/1538-3881/ab46a8}

\bibitem[{{Ormel} \& {Cuzzi}(2007)}]{Ormel_2007}
{Ormel}, C.~W., \& {Cuzzi}, J.~N. 2007, \aap, 466, 413,
  \dodoi{10.1051/0004-6361:20066899}

\bibitem[{{Paardekooper} {et~al.}(2011){Paardekooper}, {Baruteau}, \&
  {Meru}}]{Paardekooper_2011}
{Paardekooper}, S.-J., {Baruteau}, C., \& {Meru}, F. 2011, \mnras, 416, L65,
  \dodoi{10.1111/j.1745-3933.2011.01099.x}

\bibitem[{{Poppe} {et~al.}(2000){Poppe}, {Blum}, \& {Henning}}]{Poppe_2000}
{Poppe}, T., {Blum}, J., \& {Henning}, T. 2000, \apj, 533, 454,
  \dodoi{10.1086/308626}

\bibitem[{{Rafikov}(2015)}]{Rafikov_2015}
{Rafikov}, R.~R. 2015, \apj, 804, 62, \dodoi{10.1088/0004-637X/804/1/62}

\bibitem[{{Rea} {et~al.}(2024){Rea}, {Simon}, {Carrera}, {Lesur}, {Lyra},
  {Sengupta}, {Yang}, \& {Youdin}}]{Rea_2024}
{Rea}, D.~G., {Simon}, J.~B., {Carrera}, D., {et~al.} 2024, \apj, 972, 128,
  \dodoi{10.3847/1538-4357/ad57c5}

\bibitem[{{Rice}(2016)}]{Rice_2016}
{Rice}, K. 2016, \pasa, 33, e012, \dodoi{10.1017/pasa.2016.12}

\bibitem[{{Rice} {et~al.}(2004){Rice}, {Lodato}, {Pringle}, {Armitage}, \&
  {Bonnell}}]{Rice_2004}
{Rice}, W.~K.~M., {Lodato}, G., {Pringle}, J.~E., {Armitage}, P.~J., \&
  {Bonnell}, I.~A. 2004, \mnras, 355, 543,
  \dodoi{10.1111/j.1365-2966.2004.08339.x}

\bibitem[{{Ros} {et~al.}(2019){Ros}, {Johansen}, {Riipinen}, \&
  {Schlesinger}}]{Ros_2019}
{Ros}, K., {Johansen}, A., {Riipinen}, I., \& {Schlesinger}, D. 2019, \aap,
  629, A65, \dodoi{10.1051/0004-6361/201834331}

\bibitem[{{Ross} {et~al.}(2016){Ross}, {Latter}, \& {Guilet}}]{Ross_2016}
{Ross}, J., {Latter}, H.~N., \& {Guilet}, J. 2016, \mnras, 455, 526,
  \dodoi{10.1093/mnras/stv2286}

\bibitem[{{Sano} {et~al.}(2004){Sano}, {Inutsuka}, {Turner}, \&
  {Stone}}]{Sano_2004}
{Sano}, T., {Inutsuka}, S.-i., {Turner}, N.~J., \& {Stone}, J.~M. 2004, \apj,
  605, 321, \dodoi{10.1086/382184}

\bibitem[{{Sekiya} \& {Onishi}(2018)}]{Sekiya_2018}
{Sekiya}, M., \& {Onishi}, I.~K. 2018, \apj, 860, 140,
  \dodoi{10.3847/1538-4357/aac4a7}

\bibitem[{{Shadmehri} {et~al.}(2018){Shadmehri}, {Khajenabi}, {Dib}, \&
  {Inutsuka}}]{Shadmehri_2018}
{Shadmehri}, M., {Khajenabi}, F., {Dib}, S., \& {Inutsuka}, S.-i. 2018, \mnras,
  481, 5170, \dodoi{10.1093/mnras/sty2656}

\bibitem[{{Shakura} \& {Sunyaev}(1973)}]{Shakura_1973}
{Shakura}, N.~I., \& {Sunyaev}, R.~A. 1973, \aap, 24, 337

\bibitem[{{Shi} {et~al.}(2016){Shi}, {Zhu}, {Stone}, \& {Chiang}}]{Shi_2016}
{Shi}, J.-M., {Zhu}, Z., {Stone}, J.~M., \& {Chiang}, E. 2016, \mnras, 459,
  982, \dodoi{10.1093/mnras/stw692}

\bibitem[{{Shu}(1977)}]{Shu_1977}
{Shu}, F.~H. 1977, \apj, 214, 488, \dodoi{10.1086/155274}

\bibitem[{{Simon} {et~al.}(2013){Simon}, {Bai}, {Armitage}, {Stone}, \&
  {Beckwith}}]{Simon_2013}
{Simon}, J.~B., {Bai}, X.-N., {Armitage}, P.~J., {Stone}, J.~M., \& {Beckwith},
  K. 2013, \apj, 775, 73, \dodoi{10.1088/0004-637X/775/1/73}

\bibitem[{{Stammler} \& {Birnstiel}(2022)}]{Stammler_2022}
{Stammler}, S.~M., \& {Birnstiel}, T. 2022, \apj, 935, 35,
  \dodoi{10.3847/1538-4357/ac7d58}

\bibitem[{{Stammler} {et~al.}(2017){Stammler}, {Birnstiel}, {Pani{\'c}},
  {Dullemond}, \& {Dominik}}]{Stammler_2017}
{Stammler}, S.~M., {Birnstiel}, T., {Pani{\'c}}, O., {Dullemond}, C.~P., \&
  {Dominik}, C. 2017, \aap, 600, A140, \dodoi{10.1051/0004-6361/201629041}

\bibitem[{{Taki} {et~al.}(2016){Taki}, {Fujimoto}, \& {Ida}}]{Taki_2016}
{Taki}, T., {Fujimoto}, M., \& {Ida}, S. 2016, \aap, 591, A86,
  \dodoi{10.1051/0004-6361/201527732}

\bibitem[{{Toomre}(1964)}]{Toomre_1964}
{Toomre}, A. 1964, \apj, 139, 1217, \dodoi{10.1086/147861}

\bibitem[{{Wada} {et~al.}(2009){Wada}, {Tanaka}, {Suyama}, {Kimura}, \&
  {Yamamoto}}]{Wada_2009}
{Wada}, K., {Tanaka}, H., {Suyama}, T., {Kimura}, H., \& {Yamamoto}, T. 2009,
  \apj, 702, 1490, \dodoi{10.1088/0004-637X/702/2/1490}

\bibitem[{{Woitke} {et~al.}(2024){Woitke}, {Drazkowska}, {Lammer}, {Kadam}, \&
  {Marigo}}]{Woitke_2024}
{Woitke}, P., {Drazkowska}, J., {Lammer}, H., {Kadam}, K., \& {Marigo}, P.
  2024, \aap, 687, A65, \dodoi{10.1051/0004-6361/202450289}

\bibitem[{{Xu} \& {Kunz}(2021)}]{Xu_2021}
{Xu}, W., \& {Kunz}, M.~W. 2021, \mnras, 508, 2142,
  \dodoi{10.1093/mnras/stab2715}

\bibitem[{{Yang} {et~al.}(2017){Yang}, {Johansen}, \& {Carrera}}]{Yang_2017}
{Yang}, C.-C., {Johansen}, A., \& {Carrera}, D. 2017, \aap, 606, A80,
  \dodoi{10.1051/0004-6361/201630106}

\bibitem[{{Youdin} \& {Johansen}(2007)}]{Youdin_2007b}
{Youdin}, A., \& {Johansen}, A. 2007, \apj, 662, 613, \dodoi{10.1086/516729}

\bibitem[{{Youdin} \& {Goodman}(2005)}]{Youdin_2005}
{Youdin}, A.~N., \& {Goodman}, J. 2005, \apj, 620, 459, \dodoi{10.1086/426895}

\bibitem[{{Youdin} \& {Lithwick}(2007)}]{Youdin_2007}
{Youdin}, A.~N., \& {Lithwick}, Y. 2007, \icarus, 192, 588,
  \dodoi{10.1016/j.icarus.2007.07.012}

\end{thebibliography}
\bibliographystyle{aasjournal}

\end{document}